\documentclass[11pt,a4paper]{article}
\pdfoutput=1
\usepackage{jheppub}
\usepackage[T1]{fontenc}
\usepackage{microtype}
\usepackage{amsmath,amssymb,mathtools,bm}
\usepackage{booktabs,array}
\usepackage{graphicx}
\usepackage{xcolor}
\graphicspath{{figures/}}

\newcommand{\dd}{\mathrm{d}}
\newcommand{\gev}{\,\mathrm{GeV}}
\newcommand{\tev}{\,\mathrm{TeV}}
\newcommand{\fm}{\,\mathrm{fm}}
\newcommand{\Tr}{\mathrm{Tr}}
\newcommand{\ReP}{\mathrm{Re}\,}
\newcommand{\ImP}{\mathrm{Im}\,}
\newcommand{\bk}{\bm{k}}
\newcommand{\bp}{\bm{p}}
\newcommand{\bq}{\bm{q}}
\newcommand{\bP}{\bm{P}}

\newcommand{\bb}{\bm{b}}
\newcommand{\br}{\bm{r}}
\newcommand{\bu}{\bm{u}}
\newcommand{\bx}{\bm{x}}
\newcommand{\by}{\bm{y}}
\newcommand{\Qs}{Q_s}
\newcommand{\CNP}{C_{\mathrm{NP}}}
\newcommand{\Nc}{N_c}
\newcommand{\CA}{C_A}
\newcommand{\CF}{C_F}
\newcommand{\pPb}{p+\mathrm{Pb}}
\newcommand{\pp}{p+p}
\newcommand{\eAu}{e+\mathrm{Au}}
\newcommand{\ep}{e+p}
\newcommand{\DD}{D^0\bar{D}^0}
\newcommand{\BB}{B^0\bar{B}^0}
\newcommand{\FWW}{\mathcal{F}_{\mathrm{WW}}}
\newcommand{\Dlin}{\mathcal{D}_{\mathrm{lin}}}
\newcommand{\MS}{\mathbf{M}_S}

\title{\boldmath Beyond the large-$N_c$ limit in small-$x$ heavy-quark pair production}

\hypersetup{
  pdftitle={Beyond the large-Nc limit in small-x heavy-quark pair production},
  pdfauthor={Lei Wang},
  pdfsubject={JHEP manuscript},
  pdfkeywords={Color Glass Condensate, Large-Nc Expansion, Gluon Saturation, Heavy Quark Production, Transverse Momentum Resummation}
}

\author[a]{Lei Wang}
\affiliation[a]{Key Laboratory of Quark and Lepton Physics (MOE)\\
and Institute of Particle Physics, Central China Normal University,\\
No.~152 Luoyu Road, Wuhan, Hubei 430079, China}
\emailAdd{leiwang@ccnu.edu.cn}

\abstract{%
We study finite-$\Nc$ corrections to heavy-quark pair azimuthal correlations in forward proton-nucleus ($\pPb$) and electron-nucleus ($\eAu$) collisions within the Color Glass Condensate (CGC) framework combined with transverse-momentum-dependent (TMD) Sudakov resummation. In the small-$x$ CGC sector, a Duhamel expansion of the exact $2\times 2$ Gaussian evolution matrix for the fundamental quadrupole and double-dipole correlators yields closed-form coordinate-space kernels for both unpolarized and linearly polarized small-$x$ gluon TMDs at arbitrary $\Nc$. In both polarization channels, all first-derivative squared terms cancel in the sum of color-flow kernels, isolating a $+1/\Nc^2$ Weizs\"acker--Williams tail that competes with the enhanced adjoint-to-fundamental Casimir ratio ($\CA/\CF = 9/4$ versus $2$) and increases the linear polarization ratio of the effective $g+g^*\to Q\bar{Q}$ gluon TMD in the saturation regime. For soft-gluon resummation, we obtain the one-loop $3\times 3$ color-correlated soft anomalous dimension matrix for $g+g^*\to Q\bar{Q}$ from four-parton eikonal factorization, examine its massless, threshold, and 't~Hooft planar limits, and show how the non-Abelian Coulomb/Glauber phase enters at three-loop order via a double commutator away from symmetric momentum sharing. Combining both ingredients with running-coupling BK evolution, per-impact-parameter nuclear averaging, full massive off-shell improved TMD (ITMD) hard factors including linearly polarized difference terms, and off-shell $3\times 3$ hard--soft color contraction, we find that finite-$\Nc$ shifts relative to the large-$\Nc$ planar baseline cancel strongly in normalized LHCb $\pPb/\pp$ nuclear broadening ratios for all bottom-pair bins and for charm pairs with $p_T \ge 4\gev$ (and remain small in finite-angle away-side nuclear ratios across all kinematics), while both the omitted leading-$\Nc$ planar soft logarithm at low $p_T$ and modifications of individual $\pPb$ azimuthal distributions remain substantial. In contrast, heavy-quark leptoproduction at the Electron-Ion Collider involves only a one-dimensional color-octet space without singlet--octet mixing, keeping finite-$\Nc$ effects modest over the full kinematic range.
}


\begin{document}
\maketitle
\raggedbottom
\clubpenalty=10000
\widowpenalty=10000

\section{Introduction}
\label{sec:intro}

High-energy proton-nucleus ($\pPb$) collisions at the Large Hadron Collider (LHC) and deep inelastic scattering ($e+A$) at the future Electron-Ion Collider (EIC) probe nuclear matter in the highly asymmetric dilute--dense regime, where a well-calibrated projectile parton or virtual photon scatters off a dense target gluon field at small Bjorken-$x$~\cite{McLerran:1993ni,Gelis:2010nm,Morreale:2021pnn,AbdulKhalek:2021gbh,Caucal:2026yzl,Iancu:2026cgc}. As $x_A$ decreases, linear BFKL gluon branching drives up the phase-space occupancy until nonlinear gluon recombination tames the growth of the distribution and dynamically generates a semihard saturation scale $\Qs(x_A)$. Within the Color Glass Condensate (CGC) effective field theory, this high-occupancy regime is formulated in terms of lightlike Wilson lines averaged over the target color fields and evolved in rapidity via the Balitsky--Kovchegov (BK) and Jalilian-Marian--Iancu--McLerran--Weigert--Leonidov--Kovner (JIMWLK) equations~\cite{Balitsky:1995ub,Jalilian-Marian:1997jx,Kovchegov:1999yj,Iancu:2000hn,Balitsky:2006wa,Iancu:2026cgc}. Moving from qualitative signatures of saturation to quantitative precision tests in forward $p+A$ and $e+A$ scattering has required combining small-$x$ evolution systematically with higher-order perturbative matching as well as collinear, threshold, and Sudakov resummations~\cite{Mueller:2012uf,Mueller:2013wwa,Zheng:2014vka,Stasto:2018rci,Shi:2021hwx,Wang:2022zdu,Taels:2022tza,Caucal:2022ulg,Caucal:2023fsf,Caucal:2023nci,Caucal:2024bae,Caucal:2025dih}.

Among two-particle observables sensitive to transverse-momentum broadening in a saturated nuclear field, nearly back-to-back heavy-quark pairs have several practical advantages~\cite{Blaizot:2004wv,Fujii:2006ab,Dominguez:2011wm,Fujii:2013yja,Altinoluk:2021ygv}. In light-hadron pair and high-$p_T$ dijet correlations, collinear and soft Sudakov radiation from the hard scale substantially broadens the away-side peak and can obscure the nuclear saturation signal~\cite{Mueller:2013wwa,Zheng:2014vka,Caucal:2022ulg,Caucal:2023fsf,Caucal:2023nci}. For charm ($\DD$) and bottom ($\BB$) pairs, the quark mass suppresses final-state collinear radiation inside the dead cone $\theta \lesssim m_Q/E_Q$, while hard fragmentation functions retain much of the partonic kinematics. Varying $m_Q$ and the pair transverse momentum $P_T$ therefore allows one to scan across $\Qs(x_A)$~\cite{LHCb:2012aiv,LHCb:2017bdt,LHCb:2020jse,delCastillo:2020omr,Marquet:2025msx,Gao:2026msx,Salazar:2026ssz,Kang:2026ctf}.

Recent full-azimuthal CGC calculations incorporating massive-quark Sudakov resummation with the dead-cone effect~\cite{Salazar:2026ssz} predict a clear nuclear broadening of the away-side $\DD$ and $\BB$ correlation peak in forward $\pPb$ relative to $\pp$ collisions at LHCb, with saturation-scale uncertainties quoted at the few-percent level. However, like most phenomenological saturation studies to date, these calculations rely on two major approximations: multi-point Wilson-line correlators in the target are evaluated in the large-$\Nc$ limit, and Sudakov resummation retains only color-diagonal cusp and collinear logarithms. Since subleading color corrections are parametrically of order $1/\Nc^2 \approx 11\%$ at $\Nc = 3$, one must check whether neglected finite-$\Nc$ terms and color-correlated soft radiation exceed the quoted saturation uncertainty bands, or cancel sufficiently in normalized nuclear observables to allow a reliable extraction of $\Qs(x_A)$.

Addressing this question requires separating three distinct ingredients omitted from a large-$\Nc$, color-diagonal Sudakov baseline. First, in the small-$x$ nuclear Wilson-line correlators, the eikonal cross section for $g+A\to Q\bar{Q}+X$ depends on both the double-dipole correlator $S^{(2,2)}$ and the fundamental quadrupole $S^{(4)}$~\cite{Blaizot:2004wv,Fukushima:2007dy,Marquet:2010cf,Dominguez:2011wm,Iancu:2011ns,Dumitru:2011vk,Lappi:2012nh}. Going from $\Nc \to \infty$ to $\Nc = 3$ modifies the nuclear gluon distributions in two competing ways: the adjoint dipole exponent increases by $12.5\%$ as the Casimir ratio shifts from $\alpha_\infty = 2$ to $\alpha = \CA/\CF = 2\Nc^2/(\Nc^2 - 1) = 9/4$ (raising the effective adjoint saturation scale by $\alpha^{1/\gamma}$ for $L(r) \propto r^{2\gamma}$), while exact $\mathrm{SU}(\Nc)$ Fierz and Gaussian contractions generate $1/\Nc^2$-weighted Weizs\"acker--Williams (WW) quadrupole tails in both the unpolarized and linearly polarized small-$x$ gluon TMDs~\cite{Dominguez:2011gc,Iancu:2011nj,Marquet:2016cgx,Cougoulic:2026drr,Boer:2010zf,Metz:2011wb,Dominguez:2011br,Akcakaya:2012si,Pisano:2013cya,Dumitru:2015gaa,Boer:2016fqd,Marquet:2017xwy,Dumitru:2018kuw}. Second, in perturbative Sudakov resummation, wide-angle soft gluon emission in massive heavy-quark pair hadroproduction is governed by a non-diagonal $3\times 3$ soft anomalous dimension matrix $\MS(z, P_T, m_Q; \Nc)$ in color space~\cite{Kidonakis:1997gm,Kidonakis:2009ev,Mitov:2009sv,Beneke:2009rj,Ferroglia:2009ep,Ferroglia:2009ii,Zhu:2012ts,Li:2013mia,Catani:2014qha,Sun:2015doa,Catani:2018mei}. Even in the 't~Hooft large-$\Nc$ limit, this matrix does not reduce to unity: its planar limit produces an $\mathcal{O}(\Nc^0)$ single-logarithmic factor $R_{\mathrm{planar}}(b) > 1$ reflecting the reduction of the initial--final color-flow scale from the pair invariant mass squared $\hat{s}$ to $-t_1 = (1-z)\hat{s}$ and $-u_1 = z\hat{s}$. Third, subleading-color resummation effects starting at $\mathcal{O}(1/\Nc^2)$ arise from the color-singlet $Q\bar{Q}$ channel, off-diagonal singlet--octet mixing, the velocity-dependent heavy-quark dipole $\mathbf{T}_1\cdot\mathbf{T}_2$, and the non-Abelian Coulomb/Glauber imaginary phase, which does not commute with the real soft anomalous dimension away from $z = 1/2$~\cite{Forshaw:2006fk,Catani:2011st,Becher:2021zkk,Boer:2023ljq,Boer:2024xzy}.

Comparing forward heavy-quark pair hadroproduction at LHCb ($g+A \to Q\bar{Q}+X$) with leptoproduction at the EIC ($\gamma^* + A \to Q\bar{Q}+X$) provides a direct contrast in color topology. Replacing an incoming gluon by a virtual photon removes the $\mathcal{F}_{gg}^{(1,2)}$ target TMD mixing, leaving only $\FWW$, and reduces the $3\times 3$ resummation color space to a single color-octet state with no singlet--octet mixing or non-commuting Glauber phases. Analyzing both processes within the same framework therefore separates finite-$\Nc$ effects intrinsic to nuclear saturation from those tied to four-parton hadronic color flow.

To isolate perturbative and CGC finite-$\Nc$ effects without mixing in unrelated model dependencies, we adopt a well-defined theoretical setup. For the small-$x$ target correlators, we use the exact finite-$\Nc$ CGC Gaussian truncation evaluated on running-coupling BK (rcBK) dipoles rather than full lattice Langevin JIMWLK evolution; this retains exact $\mathrm{SU}(\Nc)$ Casimir algebra and four-point unitarity while providing closed-form expressions for all unpolarized and linearly polarized TMD kernels. For perturbative Sudakov evolution, we combine the one-loop $3\times 3$ soft anomalous dimension matrix with the perturbative Sudakov factor for massive quarks, off-shell $3\times 3$ hard--soft color contraction, and full massive off-shell improved TMD (ITMD) hard factors over the full azimuthal range, keeping the nonperturbative Sudakov parameterization fixed across color schemes while varying its strength and matching scale.

The paper is organized as follows. In section~\ref{sec:target}, we solve the $2\times 2$ finite-$\Nc$ CGC Gaussian evolution matrix by Duhamel expansion, derive closed-form coordinate-space kernels and $(L')^2$-cancellation sum rules for unpolarized and linearly polarized small-$x$ gluon TMDs, and evaluate them on rcBK targets. Section~\ref{sec:soft} derives the one-loop $3\times 3$ color-correlated soft anomalous dimension matrix from four-parton eikonal factorization, examines its massless, threshold, and 't~Hooft planar limits along with the three-loop double-commutator contribution of the Coulomb/Glauber phase, and contrasts this with the one-dimensional color-octet dipole in EIC leptoproduction. Section~\ref{sec:itmd} describes the full massive off-shell ITMD hard factors, off-shell $3\times 3$ hard--soft color contraction, six-scheme decomposition, smooth CSS recoil matching, systematic variations, and per-impact-parameter optical-Glauber nuclear averaging. Section~\ref{sec:results} presents the numerical results for LHCb azimuthal distributions, comparison with LHCb $\pPb$ $\DD$ data, Run~3--4 nuclear ratio predictions, away-side Gaussian broadening and finite-angle away-side ratios, and EIC heavy-quark broadening and elliptic azimuthal anisotropy $v_2(q_T)$. Our conclusions and outlook are given in section~\ref{sec:conclusions}. Technical details of the Duhamel expansion and the four-parton one-loop eikonal integrals are collected in appendices~\ref{app:duhamel} and~\ref{app:eikonal}.

\section{Finite-\texorpdfstring{$N_c$}{Nc} CGC Gaussian quadrupole and small-\texorpdfstring{$x$}{x} gluon TMDs}
\label{sec:target}

We begin with the small-$x$ nuclear Wilson-line correlators, formulating the exact finite-$\Nc$ Color Glass Condensate (CGC) Gaussian evolution of four-point operators and extracting their closed-form coordinate-space gluon TMD kernels in both the unpolarized and linearly polarized channels. This analytical baseline isolates how the enhanced adjoint Casimir ratio $\CA/\CF = 9/4$ competes with $1/\Nc^2$ quadrupole tails before soft-gluon resummation is folded in.

\subsection{Exact \texorpdfstring{$2\times 2$}{2x2} Gaussian evolution matrix}
\label{sec:gauss_matrix}

In the high-energy eikonal approximation, the interaction of a fundamental quark--antiquark dipole with a small-$x$ nuclear shockwave is encoded in the two-point Wilson-line correlator
\begin{equation}
  S^{(2)}(r) = \frac{1}{\Nc}\left\langle \Tr\!\left[V(\bx) V^\dagger(\by)\right] \right\rangle = e^{-L(r)},
  \qquad r = |\bx - \by|,
  \label{eq:S2_def}
\end{equation}
where $V(\bx) = \mathcal{P}\exp\!\left[ig\int \dd x^+\, A^-(x^+,\bx)\right]$ is a fundamental $\mathrm{SU}(\Nc)$ Wilson line and $L(r) \ge 0$. Within the CGC Gaussian (mean-field JIMWLK~\cite{Jalilian-Marian:1997jx,Iancu:2000hn}) approximation~\cite{Blaizot:2004wv,Dominguez:2011gc,Iancu:2011nj,Marquet:2016cgx,Cougoulic:2026drr}, all higher-point Wilson-line correlators are determined entirely by the two-point color-charge correlator $\Gamma(r) = \frac{2}{\CF}L(r)$ with $\CF = (\Nc^2 - 1)/(2\Nc)$.

For a four-point product $\langle V_{i_1 j_1}(\bx_1) V^\dagger_{j_2 i_2}(\bx_2) V_{i_3 j_3}(\bx_3) V^\dagger_{j_4 i_4}(\bx_4)\rangle$, color-singlet projections span a two-dimensional subspace with unnormalized basis vectors
\begin{equation}
  v_1 = \delta_{i_1 i_2}\delta_{i_3 i_4}, \qquad
  v_2 = \delta_{i_1 i_4}\delta_{i_3 i_2},
  \qquad
  \mathbf{G} = \begin{pmatrix} \Nc^2 & \Nc \\ \Nc & \Nc^2 \end{pmatrix}.
  \label{eq:singlet_basis}
\end{equation}
In this basis, the longitudinal Gaussian evolution Hamiltonian closes onto the exact $2\times 2$ matrix~\cite{Fukushima:2007dy,Dominguez:2011gc}
\begin{equation}
  \mathbf{M}(\bx_1,\bx_2,\bx_3,\bx_4) =
  \begin{pmatrix}
    -\frac{\CF}{2}(\Gamma_{12} + \Gamma_{34}) + \frac{1}{2\Nc} F_{1234} & -\frac{1}{2} F_{1432} \\
    -\frac{1}{2} F_{1234} & -\frac{\CF}{2}(\Gamma_{14} + \Gamma_{23}) + \frac{1}{2\Nc} F_{1432}
  \end{pmatrix},
  \label{eq:gauss_M}
\end{equation}
where $\Gamma_{pq} \equiv \Gamma(|\bx_p - \bx_q|)$ and the four-point exchange combinations are
\begin{equation}
  F_{1234} = \frac{1}{2}\big(\Gamma_{14} + \Gamma_{23} - \Gamma_{13} - \Gamma_{24}\big), \qquad
  F_{1432} = \frac{1}{2}\big(\Gamma_{12} + \Gamma_{34} - \Gamma_{13} - \Gamma_{24}\big).
  \label{eq:F_def}
\end{equation}
Denoting the matrix exponential by $\mathbf{E} = \exp(\mathbf{M})$, the exact finite-$\Nc$ double-dipole $S^{(2,2)}$ and fundamental quadrupole $S^{(4)}$ take the form
\begin{align}
  S^{(2,2)}(\bx_1,\bx_2;\bx_3,\bx_4) &= \frac{1}{\Nc^2}\left\langle \Tr[V_1 V_2^\dagger]\,\Tr[V_3 V_4^\dagger]\right\rangle = E_{11} + \frac{1}{\Nc} E_{21}, \label{eq:S22_exact}\\
  S^{(4)}(\bx_1,\bx_2,\bx_3,\bx_4) &= \frac{1}{\Nc}\left\langle \Tr[V_1 V_2^\dagger V_3 V_4^\dagger]\right\rangle = E_{11} + \Nc E_{21}. \label{eq:S4_exact}
\end{align}

In the coincident limit $\bx_1 = \bx_4 = \bb$ and $\bx_2 = \bx_3 = \bb'$ ($r = |\bb - \bb'|$), we have $F_{1432} = 0$ and $F_{1234} = -\Gamma(r)$, reducing eq.~\eqref{eq:gauss_M} to the lower-triangular matrix
\begin{equation}
  \mathbf{M}_0(r) = \begin{pmatrix} -\lambda & 0 \\ \lambda/\Nc & 0 \end{pmatrix},
  \qquad
  \lambda(r) \equiv \frac{\Nc}{2}\Gamma(r) = \alpha\, L(r),
  \qquad
  \alpha \equiv \frac{\CA}{\CF} = \frac{2\Nc^2}{\Nc^2 - 1}.
  \label{eq:M0_coinc}
\end{equation}
Exponentiating $\mathbf{M}_0$ enforces exact unitarity $S^{(4)}(\bb,\bb',\bb',\bb) = 1$ and yields the adjoint dipole relation
\begin{equation}
  S^{(2,2)}(\bb,\bb';\bb',\bb) = \frac{\Nc^2 - 1}{\Nc^2}\,S_A(r) + \frac{1}{\Nc^2},
  \qquad
  S_A(r) \equiv e^{-\alpha L(r)} = \big[S^{(2)}(r)\big]^{\CA/\CF}.
  \label{eq:SA_exact}
\end{equation}
Notice that for $\Nc = 3$ the Casimir ratio $\alpha = 9/4$ exceeds its large-$\Nc$ limit $\alpha_\infty = 2$ by $12.5\%$. Near the saturation boundary where the dipole exponent scales locally as $L(r) \propto r^{2\gamma}$ (with anomalous dimension $\gamma \lesssim 1$), the adjoint dipole relation $S_A(r) = e^{-\alpha L(r)}$ shifts the effective adjoint saturation scale by $Q_{s,A}^2/\Qs^2 = \alpha^{1/\gamma}$ (reducing to $Q_{s,A}^2 = \alpha\Qs^2$ in the quadratic limit $\gamma = 1$). For general non-coincident geometries, taking $\Nc \to \infty$ at fixed $L_{pq}$ reduces eq.~\eqref{eq:S4_exact} to the standard large-$\Nc$ Gaussian quadrupole~\cite{Dominguez:2011gc},
\begin{equation}
  S^{(4)}_{\Nc\to\infty}(1,2,3,4) = e^{-a} - f\,\frac{e^{-b} - e^{-a}}{a - b},
  \label{eq:S4_largeNc}
\end{equation}
with $a = L_{12}+L_{34}$, $b = L_{14}+L_{23}$, and $f = L_{14}+L_{23}-L_{13}-L_{24}$.

\subsection{Exact finite-\texorpdfstring{$N_c$}{Nc} unpolarized and linearly polarized TMD kernels}
\label{sec:tmd_theorems}

In forward $g(p_a, a) + A \to Q(\bk_1) + \bar{Q}(\bk_2) + X$ production, the eikonal scattering amplitude carries the Wilson-line structure $V(\bb + (1-z)\bu) t^a V^\dagger(\bb - z\bu) - V(\bb) t^a V^\dagger(\bb)$, where $\bu = \bx_1 - \bx_2$ is the transverse dipole separation and $\bb = z\bx_1 + (1-z)\bx_2$ is the center of energy. Expanding to linear order in $\bu$ in the back-to-back TMD regime gives
\begin{equation}
  V(\bb+(1-z)\bu) t^a V^\dagger(\bb-z\bu) - V(\bb)t^a V^\dagger(\bb)
  = u^i \left[(1-z)(\partial_i V) t^a V^\dagger - z V t^a (\partial_i V^\dagger)\right] + \mathcal{O}(\bu^2).
  \label{eq:u_expand}
\end{equation}
Squaring this amplitude and summing over the projectile gluon color index $a$ via the $\mathrm{SU}(\Nc)$ Fierz identity $t^a_{kl} t^a_{mn} = \frac{1}{2}(\delta_{kn}\delta_{lm} - \frac{1}{\Nc}\delta_{kl}\delta_{mn})$ simplifies the result: using $V(\partial_i V^\dagger) = -(\partial_i V) V^\dagger$, all four $-\frac{1}{2\Nc}$ Fierz terms carry the same sign and combine with coefficient $(1-z)^2 + z^2 + 2z(1-z) = 1$ into the Weizs\"acker--Williams correlator $\langle\Tr[(\partial_i V)V^\dagger(\bb)\,V(\partial_j V^\dagger)(\bb')]\rangle$. Averaging over the $\Nc^2 - 1$ initial gluon colors yields the exact finite-$\Nc$ back-to-back target tensor
\begin{equation}
  \mathcal{W}^{ij}(\br; z) = \frac{\Nc^2}{2(\Nc^2 - 1)} \left\{
    \big[z^2 + (1-z)^2\big]\mathcal{K}_{gg}^{(1),ij}(\br)
    + 2z(1-z)\mathcal{K}_{gg}^{(2),ij}(\br)
    - \frac{1}{\Nc^2}\mathcal{K}_{\mathrm{WW}}^{ij}(\br)
  \right\}.
  \label{eq:W_tensor_exact}
\end{equation}

For an azimuthally symmetric target, every rank-2 coordinate-space kernel decomposes uniquely into an unpolarized trace $\mathcal{K}(r)$ and a traceless linearly polarized component $\mathcal{H}(r)$,
\begin{equation}
  \mathcal{K}^{ij}(\br) = \frac{1}{2}\delta^{ij}\,\mathcal{K}(r) + \left(\hat{r}^i\hat{r}^j - \frac{1}{2}\delta^{ij}\right)\mathcal{H}(r),
  \label{eq:tensor_decomp}
\end{equation}
whose two-dimensional Fourier transform defines the momentum-space unpolarized ($f_1$) and linearly polarized ($h_1^\perp$) small-$x$ gluon TMDs (up to the universal prefactor $\Nc/(2\pi^2\alpha_s)$) through $J_0$ and $J_2$ Hankel transforms:
\begin{equation}
  f_1(k_T) = \int_0^\infty \frac{r\,\dd r}{2\pi}\, J_0(k_T r)\,\mathcal{K}(r), \qquad
  h_1^\perp(k_T) = -\int_0^\infty \frac{r\,\dd r}{2\pi}\, J_2(k_T r)\,\mathcal{H}(r),
  \label{eq:hankel_j0_j2}
\end{equation}
which must obey the Mulders--Rodrigues positivity bound $|h_1^\perp(k_T)| \le f_1(k_T)$~\cite{Mulders:2000sh,Metz:2011wb,Dominguez:2011br,Marquet:2017xwy,Wang:2026qup}.

As derived in appendix~\ref{app:duhamel} via Duhamel expansion of $\exp(\mathbf{M})$ around $\mathbf{M}_0(r)$, the first-derivative matrices of $\mathbf{M}$ commute identically with $\mathbf{M}_0$, allowing every mixed second derivative to be integrated in closed form. Introducing the Weizs\"acker--Williams multiple-scattering factor
\begin{equation}
  W(r) \equiv \frac{1 - S_A(r)}{\alpha L(r)},
  \label{eq:W_sat}
\end{equation}
together with the radial Laplacian $\nabla_r^2 L(r) = L''(r) + L'(r)/r$ and the traceless radial operator $\Dlin L(r) \equiv r\,\frac{\dd}{\dd r}(L'/r) = L''(r) - L'(r)/r$, we obtain the exact finite-$\Nc$ unpolarized TMD kernels:
\begin{align}
  \mathcal{K}_{\mathrm{dip}}(r) &= e^{-L(r)}\left[\nabla_r^2 L(r) - \big(L'(r)\big)^2\right], \label{eq:Kdip} \\
  \mathcal{K}_{\mathrm{WW}}(r) &= W(r)\,\nabla_r^2 L(r), \label{eq:Kww} \\
  \mathcal{K}_{gg}^{(1)}(r) &= \left[\left(1 - \frac{2}{\Nc^2}\right) S_A(r) + \frac{2}{\Nc^2} W(r)\right]\nabla_r^2 L(r) - \frac{\Nc^2}{\Nc^2-1}\, S_A(r)\,\big(L'(r)\big)^2, \label{eq:Kgg1} \\
  \mathcal{K}_{gg}^{(2)}(r) &= \frac{2}{\Nc^2}\Big[W(r) - S_A(r)\Big]\nabla_r^2 L(r) + \frac{\Nc^2}{\Nc^2-1}\, S_A(r)\,\big(L'(r)\big)^2. \label{eq:Kgg2}
\end{align}
Because $\hat{r}^i\hat{r}^j = \frac{1}{2}\delta^{ij} + (\hat{r}^i\hat{r}^j - \frac{1}{2}\delta^{ij})$ projects equally onto the trace and traceless channels, the exact finite-$\Nc$ \emph{linearly polarized} kernels follow simultaneously by replacing $\nabla_r^2 L(r) \to \Dlin L(r)$ while keeping all $(L')^2$ coefficients unchanged:
\begin{align}
  \mathcal{H}_{\mathrm{dip}}(r) &= e^{-L(r)}\left[\Dlin L(r) - \big(L'(r)\big)^2\right], \label{eq:Hdip} \\
  \mathcal{H}_{\mathrm{WW}}(r) &= W(r)\,\Dlin L(r), \label{eq:Hww} \\
  \mathcal{H}_{gg}^{(1)}(r) &= \left[\left(1 - \frac{2}{\Nc^2}\right) S_A(r) + \frac{2}{\Nc^2} W(r)\right]\Dlin L(r) - \frac{\Nc^2}{\Nc^2-1}\, S_A(r)\,\big(L'(r)\big)^2, \label{eq:Hgg1} \\
  \mathcal{H}_{gg}^{(2)}(r) &= \frac{2}{\Nc^2}\Big[W(r) - S_A(r)\Big]\Dlin L(r) + \frac{\Nc^2}{\Nc^2-1}\, S_A(r)\,\big(L'(r)\big)^2. \label{eq:Hgg2}
\end{align}
We have verified eqs.~\eqref{eq:Kdip}--\eqref{eq:Hgg2} numerically against direct finite-difference derivatives of $\exp(\mathbf{M})$ for multiple values of $\Nc$, finding agreement at the $10^{-8}$ level set by finite-difference truncation.

Equations~\eqref{eq:Kdip}--\eqref{eq:Hgg2} satisfy an exact sum rule in both polarization sectors: the $(L')^2$ terms cancel identically in $\mathcal{K}_{gg}^{(1)} + \mathcal{K}_{gg}^{(2)}$ and $\mathcal{H}_{gg}^{(1)} + \mathcal{H}_{gg}^{(2)}$. Substituting these kernels into eq.~\eqref{eq:W_tensor_exact}, the effective $g+g^*\to Q\bar{Q}$ gluon TMD correlators simplify to
\begin{align}
  \mathcal{W}_{\mathrm{unpol}}(r; z) = \frac{\Nc^2}{2(\Nc^2-1)} \bigg\{
    &\left[\left(z^2 + (1-z)^2 - \frac{2}{\Nc^2}\right) S_A(r) + \frac{1}{\Nc^2} W(r)\right]\nabla_r^2 L(r) \nonumber\\
    &- \frac{\Nc^2}{\Nc^2-1}(1-2z)^2 S_A(r)\big(L'(r)\big)^2
  \bigg\},
  \label{eq:Wunpol_exact}
\end{align}
with an identical expression for $\mathcal{W}_{\mathrm{lin}}(r; z)$ upon replacing $\nabla_r^2 L \to \Dlin L$.

Two immediate physical consequences follow from eq.~\eqref{eq:Wunpol_exact}. First, at symmetric momentum sharing ($z = 1/2$), the $(L')^2$ contribution vanishes identically and the effective correlator becomes a two-component superposition of the exponentially damped adjoint dipole $S_A(r)$ and the slowly decaying, power-law Weizs\"acker--Williams quadrupole tail $W(r) \sim [\alpha L(r)]^{-1}$. At $\Nc = 3$, this $+1/\Nc^2 W(r)$ tail carries a substantial fraction ($2/7 \approx 29\%$) of the weight in the dilute limit, competing directly against the $+12.5\%$ increase of the adjoint Casimir ratio in $S_A(r)$. Because $W(r) > S_A(r)$ at all $r > 0$ and overtakes $S_A(r)$ at smaller dipole separations in a heavy nucleus than in a proton, the nuclear WW quadrupole tail in isolation enhances low-$k_T$ gluons more strongly in Pb than in $p$ and therefore opposes both the larger adjoint saturation scale $Q_{s,A}^2$ and the enhanced Sudakov suppression in the color-singlet channel. Second, in the linearly polarized sector, $|\Dlin L|$ is much smaller than $\nabla_r^2 L$ at short distances (for $L \propto r^2\ln(1/r\Lambda)$, $\Dlin L$ is logarithmic-free while $\nabla_r^2 L$ is logarithmically enhanced). Whereas the dipole TMD retains its $-(L')^2$ term and saturates the positivity bound $h_1^{\perp,\mathrm{dip}}/f_1^{\mathrm{dip}} = 1$ identically, the large-$\Nc$ effective $g+g^*\to Q\bar{Q}$ linear polarization at $z=1/2$ is proportional purely to $S_A(r)\Dlin L(r)$ and is strongly suppressed inside the saturation region $k_T \lesssim \Qs$. At finite $\Nc = 3$, the surviving $+1/\Nc^2 W(r)\Dlin L(r)$ WW tail lifts this exponential suppression and nearly doubles $h_1^\perp/f_1$ for $\mathcal{W}_{gA}$ in central Pb around $k_T \sim \Qs$.

It is worth noting the scope of the finite-$\Nc$ CGC Gaussian truncation relative to full lattice Langevin JIMWLK evolution. Numerical JIMWLK benchmarks~\cite{Dumitru:2011vk,Lappi:2012nh,Alvioli:2012ba,Marquet:2016cgx,Cougoulic:2026drr,Cougoulic:2026sww,Iancu:2026cgc} have found that, when evaluated on running-coupling evolved dipoles at $\Nc = 3$, the finite-$\Nc$ Gaussian truncation reproduces four-point quadrupoles and small-$x$ gluon TMDs closely across intermediate separations $r \sim 1/\Qs$, while analytic studies beyond the Gaussian closure~\cite{Penttala:2025tmp} indicate that non-Gaussian small-$x$ evolution corrections can still appear in extended rapidities and non-symmetric configurations. Within the present study, the exact finite-$\Nc$ Gaussian truncation isolates the $\mathrm{SU}(\Nc)$ Casimir and Fierz structures analytically while keeping the small-$x$ dipole input directly comparable to rcBK baselines. Whereas ref.~\cite{Cougoulic:2026drr} recently compiled general-$\mathrm{SU}(\Nc)$ unpolarized TMD expressions in the Gaussian initial condition, eqs.~\eqref{eq:Kdip}--\eqref{eq:Wunpol_exact} extend this closed-form structure simultaneously to the linearly polarized kernels $\mathcal{H}_{gg}^{(1,2)}$ and establish the exact $(L')^2$-cancellation sum rules (together with the exact identity $\mathcal{K}_{gg}^{(1)} - \mathcal{K}_{gg}^{(2)} = \mathcal{K}_{\mathrm{ADP}}$ and $\mathcal{H}_{\mathrm{ADP}} = \mathcal{F}_{\mathrm{ADP}}$ at arbitrary $\Nc$) needed for our coupled CGC and Sudakov-resummation analysis below.

\begin{figure}[tbp]
  \centering
  \includegraphics[width=\textwidth]{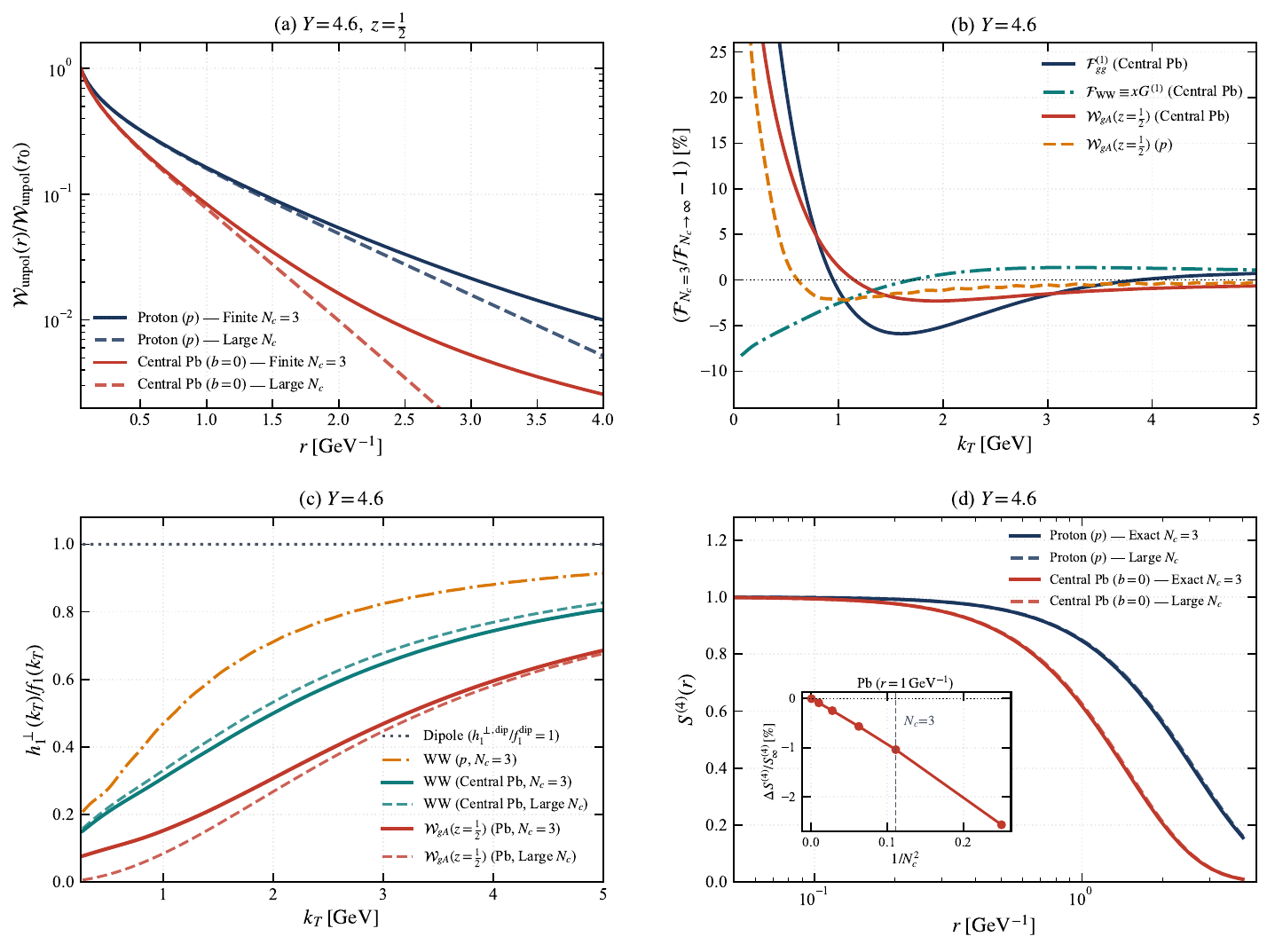}
  \caption{Finite-$\Nc$ CGC Gaussian gluon TMDs and quadrupole correlator evaluated on AAMQS rcBK dipoles at $Y = 4.6$ ($x_A = 10^{-4}$): (a) coordinate-space effective kernel $\mathcal{W}_{\mathrm{unpol}}(r; z=1/2)$ normalized at $r_0 = 0.05\gev^{-1}$, showing the $+1/\Nc^2$ WW tail at large $r$; (b) momentum-space unpolarized finite-$\Nc$ shifts in $\mathcal{F}_{gg}^{(1)}$, $\FWW$, and $\mathcal{W}_{gA}$; (c) linear polarization ratio $h_1^\perp(k_T)/f_1(k_T)$ contrasting exact dipole saturation ($=1$), WW suppression, and the finite-$\Nc$ enhancement of $\mathcal{W}_{gA}$ in central Pb; (d) exact finite-$\Nc$ vs.\ large-$\Nc$ quadrupole $S^{(4)}(r)$ in the square configuration, with inset illustrating linear $1/\Nc^2$ scaling.}
  \label{fig:tmds}
\end{figure}

To evaluate these kernels on evolved nuclear targets, we use the Balitsky running-coupling BK (rcBK) AAMQS proton and lead dipole amplitudes~\cite{Balitsky:1995ub,Kovchegov:1999yj,Balitsky:2006wa,Albacete:2010sy,Lappi:2013zma} at $Y = 4.6$ ($x_A = 10^{-4}$), shown in figure~\ref{fig:tmds}. To avoid spline oscillations in the deep-saturation region ($1 - N(Y, r) \to 0$) and to smooth the second-derivative corner introduced by the infrared freezing cutoff $\alpha_s(r) \le 0.7$ of the AAMQS rcBK grid near $r \simeq 8.3\gev^{-1}$, we fit a quintic smoothing spline ($s = 3\times 10^{-4}$) to $g(u) = \ln L(u)$ versus $u = \ln r$ on $10^{-13} < N < 1 - 10^{-5}$ and compute $L' = (L/r)g'$, $\nabla_r^2 L = (L/r^2)[g''+(g')^2]$, and $\Dlin L = (L/r^2)[g''+(g')^2 - 2g']$. The $J_0$ and $J_2$ Hankel transforms are performed by sub-interval Gauss--Legendre quadrature up to $r_{\max} = 48\gev^{-1}$ with an infrared regulator $\exp[-(r/r_{\mathrm{IR}})^4]$ at $r_{\mathrm{IR}} = 32\gev^{-1}$ applied to the power-law WW and $gg^{(1,2)}$ kernels. The exponentially damped dipole kernel is left unregulated, preserving $h_1^{\perp,\mathrm{dip}}/f_1^{\mathrm{dip}} = 1$ to $4\times 10^{-6}$.

Figure~\ref{fig:tmds} illustrates how these coordinate-space structures carry over into momentum space. In panel~(a), the finite-$\Nc$ effective kernel $\mathcal{W}_{\mathrm{unpol}}(r; z=1/2)$ rises above its exponentially damped large-$\Nc$ counterpart for $r \gtrsim 1\gev^{-1}$ in central Pb ($\Qs^2 \simeq 1.8\gev^2$) due to the slowly decaying $+1/\Nc^2 W(r)$ quadrupole tail; in the more dilute proton, this crossover occurs at larger $r$. After the $J_0$ transform (panel~b), the large-$r$ tail produces an excess of low-$k_T$ gluons at $k_T \lesssim \Qs$ and a mild intermediate-$k_T$ depletion from the larger adjoint Casimir ratio $\alpha = 9/4$, before all distributions approach the same perturbative high-$k_T$ limit. The pure WW distribution $\FWW$ shows the opposite low-$k_T$ behavior, a modest suppression at finite $\Nc$, because it has no compensating $+1/\Nc^2$ tail relative to $S_A$ and responds only to $\alpha = 9/4 > 2$. In the linearly polarized sector (panel~c), the dipole TMD saturates the Mulders--Rodrigues bound $h_1^{\perp,\mathrm{dip}}/f_1^{\mathrm{dip}} = 1$ identically and pure WW polarization is slightly lowered by the larger adjoint Casimir ratio, whereas the $+1/\Nc^2 W(r)\Dlin L(r)$ tail lifts the exponential saturation suppression of $\mathcal{W}_{gA}$, nearly doubling $h_1^\perp/f_1$ in central Pb around $k_T \sim 1\text{--}2\gev$. Finally, panel~(d) compares the exact square quadrupole $S^{(4)}(r)$ with its large-$\Nc$ limit~\eqref{eq:S4_largeNc}: the difference stays at the percent level for $\Nc = 3$ and scales linearly with $1/\Nc^2$ (inset), confirming that the larger finite-$\Nc$ modifications of the $g+A \to Q\bar{Q}$ gluon TMDs arise from the Fierz projections and radial derivatives in eq.~\eqref{eq:Wunpol_exact} rather than from a large distortion of $S^{(4)}$ itself.

\section{Color-correlated \texorpdfstring{$3\times 3$}{3x3} soft anomalous dimension matrix and Sudakov resummation}
\label{sec:soft}

Restoring finite-$\Nc$ dynamics solely inside the target shockwave correlators would yield an unbalanced color-space treatment unless the wide-angle soft gluon radiation accompanying the colored $g+g^* \to Q\bar{Q}$ hard process is resummed at matching color accuracy. In contrast to color-singlet production, where Sudakov resummation reduces to a single scalar exponent, heavy-quark pair production involves multiple color-flow topologies whose soft emissions interfere non-trivially across both leading-$\Nc$ and $1/\Nc^2$ orders.

\subsection{Color space and first-principles four-parton eikonal factorization}
\label{sec:eikonal_MS}

In the orthonormal $s$-channel color basis for $g(p_a, a) + g^*(p_b, b) \to Q(p_1, i) + \bar{Q}(p_2, j)$ at $\Nc \ge 3$~\cite{Kidonakis:1997gm,Zhu:2012ts,Li:2013mia} (for $\Nc = 2$, $d^{abc} = 0$ and the basis reduces to $\{|1\rangle, |8_A\rangle\}$),
\begin{equation}
  \begin{gathered}
    |1\rangle = \frac{\delta^{ab}\delta_{ij}}{\sqrt{\Nc(\Nc^2-1)}}, \\[4pt]
    |8_S\rangle = \sqrt{\frac{2\Nc}{(\Nc^2-1)(\Nc^2-4)}}\,d^{abc}t^c_{ij}, \qquad
    |8_A\rangle = \sqrt{\frac{2}{\Nc(\Nc^2-1)}}\,i f^{abc}t^c_{ij},
  \end{gathered}
  \label{eq:color_basis}
\end{equation}
the tree-level back-to-back color amplitude $(1-z)\,t^b t^a + z\,t^a t^b$ projects onto the real vector
\begin{equation}
  |\mathcal{M}_0(z)\rangle = \begin{pmatrix}
    \frac{1}{2}\sqrt{\frac{\Nc^2 - 1}{\Nc}}, &
    \frac{1}{2}\sqrt{\frac{(\Nc^2 - 1)(\Nc^2 - 4)}{2\Nc}}, &
    -\left(\frac{1}{2} - z\right)\sqrt{\frac{\Nc(\Nc^2 - 1)}{2}}
  \end{pmatrix}^T,
  \label{eq:M0_vec}
\end{equation}
whose squared norm $\frac{1}{\Nc^2-1}\langle\mathcal{M}_0|\mathcal{M}_0\rangle = \frac{\Nc}{4}[z^2 + (1-z)^2 - 1/\Nc^2]$ matches the exact finite-$\Nc$ dilute coefficient of eq.~\eqref{eq:Wunpol_exact}. In the all-outgoing color-charge convention ($\mathbf{T}_a + \mathbf{T}_b + \mathbf{T}_1 + \mathbf{T}_2 = 0$), the diagonal dipole color operators are
\begin{equation}
  \mathbf{T}_1\cdot\mathbf{T}_2 = \mathrm{diag}\!\left(-\CF,\,\frac{1}{2\Nc},\,\frac{1}{2\Nc}\right), \qquad
  \mathbf{T}_a\cdot\mathbf{T}_b = \mathrm{diag}\!\left(-\Nc,\,-\frac{\Nc}{2},\,-\frac{\Nc}{2}\right),
  \label{eq:T12_Tab}
\end{equation}
while the initial--final combinations take the symmetric $3\times 3$ forms
\begin{equation}
  \begin{aligned}
    \mathbf{T}_a\cdot\mathbf{T}_1 + \mathbf{T}_b\cdot\mathbf{T}_2 &=
    \begin{pmatrix} 0 & 0 & -\sqrt{2} \\ 0 & -\frac{\Nc}{2} & -\frac{\sqrt{\Nc^2-4}}{2} \\ -\sqrt{2} & -\frac{\sqrt{\Nc^2-4}}{2} & -\frac{\Nc}{2} \end{pmatrix}, \\[4pt]
    \mathbf{T}_a\cdot\mathbf{T}_2 + \mathbf{T}_b\cdot\mathbf{T}_1 &=
    \begin{pmatrix} 0 & 0 & \sqrt{2} \\ 0 & -\frac{\Nc}{2} & \frac{\sqrt{\Nc^2-4}}{2} \\ \sqrt{2} & \frac{\sqrt{\Nc^2-4}}{2} & -\frac{\Nc}{2} \end{pmatrix}.
  \end{aligned}
  \label{eq:Ta1_Ta2}
\end{equation}

Within SCET/bHQET transverse-momentum resummation for heavy-quark pairs~\cite{Mitov:2009sv,Becher:2009kw,Beneke:2009rj,Ferroglia:2009ep,Ferroglia:2009ii,Zhu:2012ts,Li:2013mia,Catani:2014qha}, evolving the hard amplitude from $\mu = \sqrt{\hat{s}}$ down to the smooth CSS evolution scale $\mu_b(b, \mu) = c_0 / \sqrt{b_*^2 + (c_0/\mu)^2}$ (with $b_* = b/\sqrt{1 + b^2/b_{\max}^2}$ and $c_0 = 2e^{-\gamma_E}$, satisfying $\mu_b(0,\mu) = \mu$, $\dd\mu_b/\dd b|_{b=0} = 0$, and $\mu_b \le \mu$ everywhere) yields
\begin{equation}
  |\mathcal{M}(\mu_b)\rangle = e^{-\frac{1}{2} S_P(b;\mu,m_Q)} \mathbf{U}(\mu_b, \mu)\,|\mathcal{M}(\mu)\rangle,
  \qquad
  \mathbf{U}(\mu_b, \mu) = \exp\!\left[-\frac{1}{2} I_{\alpha_s}(\mu, \mu_b)\,\MS\right],
  \label{eq:RG_evol}
\end{equation}
where $I_{\alpha_s}(\mu, \mu_b) \equiv \int_{\mu_b^2}^{\mu^2} \frac{\dd\bar{\mu}^2}{\bar{\mu}^2} \frac{\alpha_s(\bar{\mu})}{4\pi}$, and the color-diagonal perturbative Sudakov exponent for massive quarks matches ref.~\cite{Salazar:2026ssz}:
\begin{align}
  S_P(b; \mu, m_Q) = \int_{\mu_b^2}^{\mu^2} \frac{\dd\bar{\mu}^2}{\bar{\mu}^2} \frac{\alpha_s(\bar{\mu})}{2\pi} \bigg\{
    &(2\CA + 2\CF)\ln\frac{\mu^2}{\bar{\mu}^2} - 2\beta_0(n_f) - 3\CF \nonumber\\
    &+ \Theta(m_Q^2 - \bar{\mu}^2)\left[\CF - 2\CF\ln\frac{m_Q^2}{\bar{\mu}^2}\right]
  \bigg\},
  \label{eq:SP_def}
\end{align}
with $\beta_0(n_f) = (11\CA - 2n_f)/12$.

To fix every sign and prefactor in the scale-independent color-correlated soft anomalous dimension matrix $\MS$, we evaluate the one-loop four-parton real eikonal radiation spectrum $-\sum_{I,J}(\mathbf{T}_I\cdot\mathbf{T}_J)I_{IJ}$ at fixed soft transverse momentum $k_\perp \ll m_Q, P_T$ in appendix~\ref{app:eikonal}. Using color conservation to separate the diagonal cusp and massive collinear logarithms already contained in eq.~\eqref{eq:SP_def}, we obtain (in $\alpha_s/(4\pi)$ units)
\begin{align}
  \MS(z, P_T, m_Q; \Nc) &= -4(\mathbf{T}_a\cdot\mathbf{T}_1 + \mathbf{T}_b\cdot\mathbf{T}_2)\ln(1-z)
  - 4(\mathbf{T}_a\cdot\mathbf{T}_2 + \mathbf{T}_b\cdot\mathbf{T}_1)\ln z \nonumber\\
  &\quad - 4(\mathbf{T}_1\cdot\mathbf{T}_2)\,f(\beta)
  + 4 i\pi\,(\mathbf{T}_a\cdot\mathbf{T}_b),
  \label{eq:MS_exact}
\end{align}
where $\hat{s} = (P_T^2 + m_Q^2)/[z(1-z)]$, $\beta = \sqrt{1 - 4m_Q^2/\hat{s}}$ is the heavy-quark center-of-mass velocity, and the IR-finite velocity function is
\begin{equation}
  f(\beta) \equiv \frac{1+\beta^2}{2\beta}\left(\ln\frac{1+\beta}{1-\beta} - i\pi\right) - \ln\frac{\hat{s}}{m_Q^2}.
  \label{eq:fbeta}
\end{equation}
The color-correlated soft factor multiplying $e^{-S_P(b)}$ in impact-parameter space in the back-to-back limit is then
\begin{equation}
  R_{\mathrm{full}}(b) \equiv \frac{\langle\mathcal{M}_0(z) | \mathbf{U}^\dagger(\mu_b,\mu)\mathbf{U}(\mu_b,\mu) | \mathcal{M}_0(z)\rangle}{\langle\mathcal{M}_0(z) | \mathcal{M}_0(z)\rangle},
  \label{eq:Rfull_def}
\end{equation}
with $\mathbf{U}(\mu_b, \mu) = \exp\!\left[-\frac{1}{2} I_{\alpha_s}(\mu, \mu_b)\,\MS\right]$. Consistent with heavy-quark SCET/bHQET renormalization-group evolution~\cite{Ferroglia:2009ep,Li:2013mia,Catani:2014qha}, the velocity-dependent dipole function $f(\beta)$ in $\MS$ depends on the external heavy-quark kinematics and is evolved continuously over $[\mu_b, \mu]$ without an artificial step switch at $\bar{\mu} = m_Q$, while smoothly vanishing as $m_Q \to 0$. We have verified eq.~\eqref{eq:MS_exact} directly against numerical integration of the four-parton one-loop eikonal phase space across all entries of the $3\times 3$ matrix to better than $10^{-11}$.

The overall sign of the angular terms in eq.~\eqref{eq:MS_exact} carries a clear physical meaning. Because $\mathbf{T}_a\cdot\mathbf{T}_1 + \mathbf{T}_b\cdot\mathbf{T}_2$ has eigenvalues $\{0, -\Nc\}$ in the octet sector while $\ln(1-z)$ and $\ln z$ are negative, the real octet eigenvalues of $\MS$ are negative, giving $R_{\mathrm{soft}}(b) > 1$ relative to $S_P(b; \mu=\sqrt{\hat{s}})$. Physically, the initial--final planar dipoles radiate at the smaller momentum-transfer scales $-t_1 = (1-z)\hat{s} < \hat{s}$ and $-u_1 = z\hat{s} < \hat{s}$, so using $\mu^2 = \hat{s}$ in the diagonal Sudakov factor overestimates perturbative damping unless compensated by $\MS$.

\subsection{Exact physical limits and the three-loop Coulomb/Glauber phase}
\label{sec:MS_limits}

The matrix $\MS(z, P_T, m_Q; \Nc)$ obeys four instructive analytical limits that clarify the interplay between leading-$\Nc$ and $1/\Nc^2$ soft radiation:

\paragraph{Massless limit ($m_Q \to 0$, $\beta \to 1$).}
In the relativistic limit, $\ln[(1+\beta)/(1-\beta)] = \ln[(1+\beta)^2/(4m_Q^2/\hat{s})] \to \ln(\hat{s}/m_Q^2)$, so $\ReP f(\beta) = \mathcal{O}(m_Q^2/\hat{s}) \to 0$ identically. All $\mathbf{T}_1\cdot\mathbf{T}_2\ln m_Q$ collinear mass logarithms cancel, smoothly recovering the massless $gg \to q\bar{q}$ soft anomalous dimension matrix.

\paragraph{Threshold limit ($\beta \to 0$, $z = 1/2$, $\hat{s} = 4m_Q^2$).}
All off-diagonal entries proportional to $\ln[(1-z)/z]$ vanish and $\MS$ diagonalizes in $\{|1\rangle, |8_S\rangle, |8_A\rangle\}$. Combining $\ReP\MS$ with the static heavy-quark constant $S_P^{\mathrm{const}} = 4\CF\ln(\hat{s}/m_Q^2) - 4\CF = 8\CF\ln 2 - 4\CF$ inside eq.~\eqref{eq:SP_def} yields
\begin{equation}
  S_P^{\mathrm{const}} + \ReP M_{11}\big|_{\beta=0} = 0, \qquad
  \ReP\big(M_{88} - M_{11}\big)\big|_{\beta=0} = -2\CA.
  \label{eq:threshold_exact}
\end{equation}
The physical meaning is straightforward: a static color-singlet $Q\bar{Q}$ pair at rest carries no net color charge and emits no soft gluons (leaving only the initial $gg$ dipole radiation $4\CA L_s$), whereas a static color-octet pair acts as a heavy adjoint source whose interference with the incoming gluons reduces the central-rapidity radiation density from $4\CA$ to $4\CA - \CA/\cosh^2 y$, integrating to $-2\CA$.

\paragraph{'t~Hooft large-$\Nc$ planar limit ($\Nc \to \infty$ at fixed $\Nc I_{\alpha_s}$).}
Both the octet heavy-quark dipole $\mathbf{T}_1\cdot\mathbf{T}_2 = 1/(2\Nc)$ and the singlet--octet mixing entries vanish as $\mathcal{O}(1/\Nc^2)$, collapsing eq.~\eqref{eq:Rfull_def} onto the decoupled planar soft factor
\begin{equation}
  R_{\mathrm{planar}}(b) = \frac{z^2\, e^{-4\Nc\ln(1-z)\,I_{\alpha_s}} + (1-z)^2\, e^{-4\Nc\ln(z)\,I_{\alpha_s}}}{z^2 + (1-z)^2} + \mathcal{O}\!\left(\frac{1}{\Nc^2}\right).
  \label{eq:Rplanar_exact}
\end{equation}
Importantly, $R_{\mathrm{planar}}(b) > 1$ is a \emph{leading-$\Nc$} ($\mathcal{O}(\Nc^0)$) correction relative to the color-diagonal Sudakov baseline $R = 1$ of ref.~\cite{Salazar:2026ssz}. Only the ratio $R_{\mathrm{full}}(b)/R_{\mathrm{planar}}(b)$ isolates genuine $\mathcal{O}(1/\Nc^2)$ resummation effects. Because the $1/\Nc^2$-weighted color-singlet channel $|1\rangle$ has zero initial--final color flow, it does not benefit from the scale reduction $\hat{s} \to -t_1, -u_1$ and therefore undergoes stronger Sudakov suppression than the octet flows, pushing $R_{\mathrm{full}}(b) < R_{\mathrm{planar}}(b)$ by roughly $10\%\text{--}15\%$ at $b \sim 1\gev^{-1}$.

\paragraph{Non-Abelian Coulomb/Glauber phase at asymmetric $z \neq 1/2$.}
Writing $\MS = \mathbf{A} + i\mathbf{B}$ with real symmetric $\mathbf{A} = \ReP\MS$ and $\mathbf{B} = \ImP\MS = 4\pi(\mathbf{T}_a\cdot\mathbf{T}_b) + 4\pi(\mathbf{T}_1\cdot\mathbf{T}_2)\frac{1+\beta^2}{2\beta}$, we observe that at $z = 1/2$ both $\mathbf{A}$ and $\mathbf{B}$ are diagonal, so $[\mathbf{A},\mathbf{B}] = 0$ and the imaginary phase cancels out of $\mathbf{U}^\dagger\mathbf{U}$ to all orders. Away from $z = 1/2$, however, $|1\rangle\leftrightarrow|8_A\rangle$ mixing in $\mathbf{A}$ fails to commute with the singlet--octet phase splitting in $\mathbf{B}$. Expanding $\mathbf{U}^\dagger\mathbf{U} = e^{-t(\mathbf{A}-i\mathbf{B})}e^{-t(\mathbf{A}+i\mathbf{B})}$ ($t = \frac{1}{2}I_{\alpha_s}$) via the Baker--Campbell--Hausdorff (BCH) formula shows that $\mathbf{B}$ cancels identically at one loop ($\mathcal{O}(I_{\alpha_s})$) and at two loops ($\mathcal{O}(I_{\alpha_s}^2)$, since $i[\mathbf{A},\mathbf{B}]$ is antisymmetric and vanishes in any real state $|\mathcal{M}_0\rangle$), entering first at \emph{three-loop order} via the double commutator:
\begin{equation}
  R_{\mathrm{full}}(b) - R_{\mathrm{full}}^{\mathrm{no\,Im}}(b)
  = -\frac{I_{\alpha_s}^3}{24}\,\frac{\langle\mathcal{M}_0 | [\mathbf{B}, [\mathbf{A}, \mathbf{B}]] | \mathcal{M}_0\rangle}{\langle\mathcal{M}_0 | \mathcal{M}_0\rangle} + \mathcal{O}(I_{\alpha_s}^4).
  \label{eq:BCH_3loop}
\end{equation}
This is the heavy-quark TMD analogue of the Coulomb/Glauber phase obstruction well known from super-leading logarithms~\cite{Forshaw:2006fk,Catani:2011st,Becher:2021zkk,Boer:2023ljq,Boer:2024xzy}. Because $\mathbf{B}^2$ carries a $(4\pi)^2$ enhancement, this formally three-loop effect produces a noticeable few-percent suppression of $R_{\mathrm{full}}(b)$ at asymmetric kinematics such as $z \approx 0.25$ and $b \sim 1\gev^{-1}$.

\subsection{Contrast with the EIC 1D color-octet soft dipole}
\label{sec:EIC_soft}

In heavy-quark pair leptoproduction at the EIC ($\gamma^*(Q^2) + g^*(k_T) \to Q\bar{Q}$), the virtual photon is colorless ($\mathbf{T}_a = 0$), collapsing the color space to a single octet state with $\mathbf{T}_b\cdot\mathbf{T}_1 = \mathbf{T}_b\cdot\mathbf{T}_2 = -\CA/2$ and a repulsive heavy-quark dipole $\mathbf{T}_1\cdot\mathbf{T}_2 = +1/(2\Nc)$. There is neither singlet--octet mixing nor a non-commuting Coulomb/Glauber phase, and the one-loop soft factor reduces to the scalar exponential
\begin{equation}
  R_{\mathrm{EIC}}(b) = \exp\!\left\{ \left[-2\Nc\ln\big(z(1-z)\big) + \frac{2}{\Nc}\,\ReP f(\beta)\right] I_{\alpha_s}(\mu, \mu_b) \right\}.
  \label{eq:REIC_exact}
\end{equation}
Since $\ReP f(\beta) < 0$ for all $\beta < 1$, the sole subleading-color term $-\frac{2}{\Nc}\ReP f(\beta) > 0$ slightly enhances soft radiation ($R_{\mathrm{EIC}}^{\mathrm{fin}} < R_{\mathrm{EIC}}^{\mathrm{lnc}}$) and vanishes smoothly in both the massless ($\beta \to 1$) and large-$\Nc$ limits.

\begin{figure}[tbp]
  \centering
  \includegraphics[width=\textwidth]{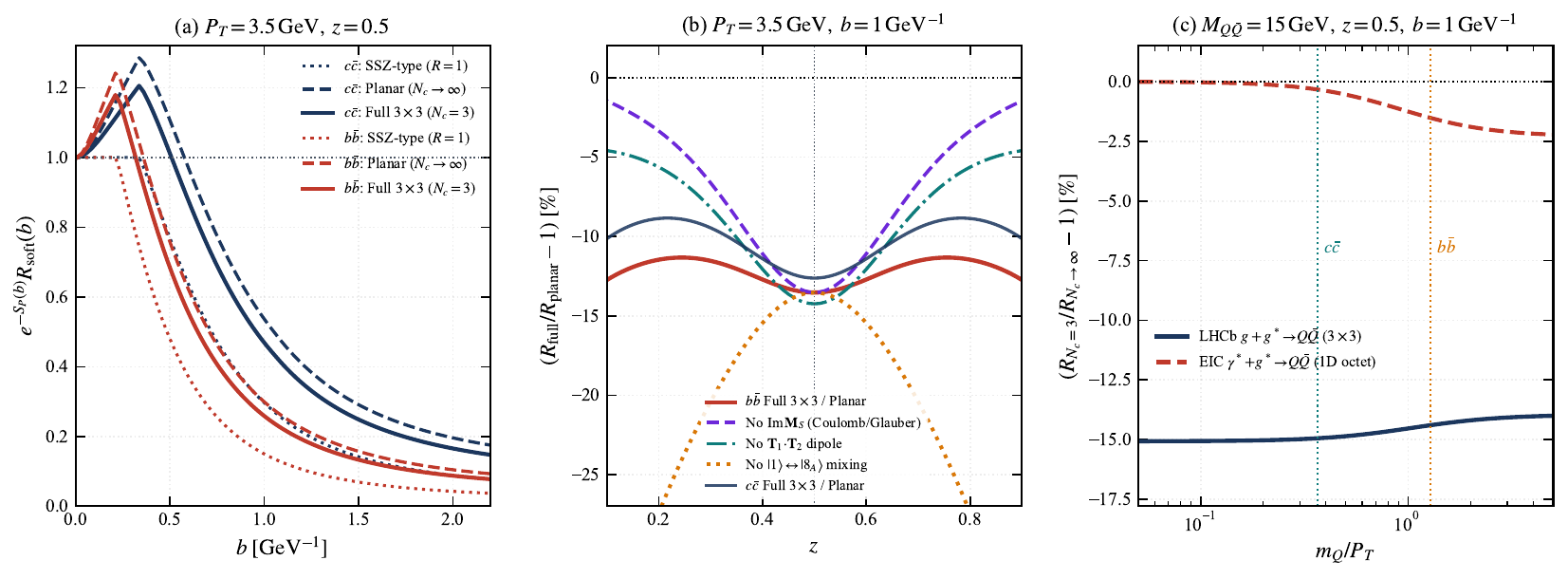}
  \caption{Evolution of the $3\times 3$ color-correlated soft anomalous dimension matrix: (a) combined perturbative kernel $e^{-S_P(b)}R_{\mathrm{soft}}(b)$ at $P_T = 3.5\gev, z=0.5$ separating the leading-$\Nc$ planar scale shift ($R_{\mathrm{planar}} > 1$) from the finite-$\Nc$ color-singlet suppression ($R_{\mathrm{full}} < R_{\mathrm{planar}}$); (b) four-way decomposition of $(R_{\mathrm{full}}/R_{\mathrm{planar}} - 1)$ vs.\ $z$ at $b = 1.0\gev^{-1}$, exposing the exact three-loop Coulomb/Glauber phase effect away from $z=1/2$; (c) mass-ratio $m_Q/P_T$ scan at fixed $M_{Q\bar{Q}} = 15\gev$ contrasting LHCb ($3\times 3$ matrix) against EIC (1D octet $\mathbf{T}_1\cdot\mathbf{T}_2$ dipole).}
  \label{fig:soft}
\end{figure}

Figure~\ref{fig:soft} summarizes the numerical size of these soft-matrix effects as a function of $b$, $z$, and $m_Q/P_T$. In panel~(a), replacing the color-diagonal SSZ-type baseline $R = 1$ by the leading-$\Nc$ planar factor $R_{\mathrm{planar}}$ raises the perturbative kernel $e^{-S_P(b)}R_{\mathrm{soft}}(b)$ appreciably at $b \sim 0.3\text{--}1.5\gev^{-1}$ (narrowing the perturbative recoil distribution), while restoring the full $3\times 3$ matrix lowers it again because the $1/\Nc^2$-weighted color-singlet component experiences stronger Sudakov suppression. Since both $|1\rangle\leftrightarrow|8_A\rangle$ mixing and the Coulomb/Glauber commutator vanish at $z = 1/2$, symmetric kinematics probes only the diagonal singlet suppression ($-12.6\%$ for $c\bar{c}$ and $-13.5\%$ for $b\bar{b}$ at $b = 1\gev^{-1}$). Moving away from $z = 1/2$ at $b = 1\gev^{-1}$ (panel~b), switching off $|1\rangle\leftrightarrow|8_A\rangle$ mixing deepens the singlet suppression steeply toward small or large $z$, whereas restoring mixing partially offsets this drop and turns on the exact three-loop Coulomb/Glauber phase~\eqref{eq:BCH_3loop} (contributing $-4.5\%$ for $c\bar{c}$ and $-6.8\%$ for $b\bar{b}$ at $z = 0.25$), leaving a nearly flat net finite-$\Nc$ suppression of $9\%\text{--}13.5\%$ across $z$. Finally, scanning $m_Q/P_T$ at fixed $M_{Q\bar{Q}} = 15\gev$ (panel~c) compares hadroproduction and leptoproduction directly: the $3\times 3$ LHCb soft matrix retains a $14\%\text{--}15\%$ finite-$\Nc$ suppression from the massless limit to threshold, whereas the EIC 1D octet dipole vanishes smoothly as $m_Q/P_T \to 0$ and remains modest ($-0.4\%$ for $c\bar{c}$, $-1.6\%$ for $b\bar{b}$, and $\lesssim 2.3\%$ near threshold).

\section{Full-angle massive ITMD matching and nuclear inputs}
\label{sec:itmd}

The analytical results of sections~\ref{sec:target} and~\ref{sec:soft} fix the target gluon TMDs and soft anomalous dimension matrix in the back-to-back regime ($\delta \equiv \pi - \Delta\Phi \to 0$). Realistic collider measurements at LHCb, however, integrate over the entire azimuthal range $\Delta\Phi \in [0,\pi]$, where the pair transverse momentum imbalance $\bk_T = \bk_1 + \bk_2$ becomes comparable to the hard scale $P_T$ and strict TMD factorization receives large kinematic-twist corrections. To connect our finite-$\Nc$ formalism smoothly from the back-to-back peak to the collinear near side ($\delta \to \pi$), we couple the exact finite-$\Nc$ Gaussian gluon TMDs and $3\times 3$ soft matrix to the full gauge-invariant off-shell massive ITMD factorization formulas of Altinoluk, Marquet, and Taels~\cite{Kotko:2015ura,vanHameren:2016ftb,Altinoluk:2019fui,Fujii:2020bkl,Altinoluk:2021ygv}.

In $g(p_a) + g^*(\bk_T) \to Q(\bk_1) + \bar{Q}(\bk_2)$ ($\bk_T = \bk_1+\bk_2$, $\bP_T = (1-z)\bk_1 - z\bk_2$), the two color-ordered amplitudes built from the massive light-front wavefunctions $\bm{\Psi}(\bp) = \bp/(\bp^2+m_Q^2)$ and $\Phi(\bp) = m_Q/(\bp^2+m_Q^2)$ are
\begin{align}
  \mathcal{A}_1\;(t^b t^a): &\quad \bm{H}_1 = \bm{\Psi}(-\bk_2) - \bm{\Psi}(\bP_T), \qquad \Delta\Phi_1 = \Phi(\bk_2) - \Phi(\bP_T), \nonumber\\
  \mathcal{A}_2\;(t^a t^b): &\quad \bm{H}_2 = \bm{\Psi}(\bP_T) - \bm{\Psi}(\bk_1), \qquad\;\; \Delta\Phi_2 = \Phi(\bP_T) - \Phi(\bk_1),
  \label{eq:itmd_amps}
\end{align}
yielding the non-sense off-shell hard kernels
\begin{equation}
  K_{gg}^{(1)} = \frac{|\mathcal{A}_1|^2 + |\mathcal{A}_2|^2}{\bk_T^2}, \qquad
  K_{gg}^{(2)} = \frac{2\,\ReP(\mathcal{A}_1\mathcal{A}_2^*)}{\bk_T^2}, \qquad
  K_{\mathrm{WW}} = K_{gg}^{(1)} + K_{gg}^{(2)},
  \label{eq:itmd_K}
\end{equation}
where $|\mathcal{A}_r|^2 = [z^2+(1-z)^2]\bm{H}_r^2 + (\Delta\Phi_r)^2$, matching $H_1^{ns} = K_{\mathrm{WW}}$ and $H_{\mathrm{ADP}}^{ns} = -K_{gg}^{(2)}$ in ref.~\cite{Altinoluk:2021ygv}. For massive quarks, however, the full ITMD cross section also includes the gauge-invariant linearly polarized difference hard factor $H_1^h(\bk_1,\bk_2)$ multiplying $(\mathcal{H}-\mathcal{F})$ (whereas the adjoint-dipole channel obeys $\mathcal{H}_{\mathrm{ADP}} = \mathcal{F}_{\mathrm{ADP}}$ identically for all $\Nc$ and needs no $H_{\mathrm{ADP}}^h$ term)~\cite{Altinoluk:2021ygv}:
\begin{align}
  \mathcal{I}_{\mathrm{ITMD}}^{gA}(\tilde{\bk}_1,\tilde{\bk}_2) &= \frac{\Nc^2}{2(\Nc^2-1)}\bigg\{
    K_{gg}^{(1)}\mathcal{F}_{gg}^{(1)} + K_{gg}^{(2)}\mathcal{F}_{gg}^{(2)} - \frac{1}{\Nc^2}K_{\mathrm{WW}}\FWW \nonumber\\
    &\qquad\qquad\qquad + H_1^h\left[\big(\mathcal{H}_{gg}^{(1)} - \mathcal{F}_{gg}^{(1)}\big) - \frac{1}{\Nc^2}\big(\mathcal{H}_{\mathrm{WW}} - \FWW\big)\right]
  \bigg\},
  \label{eq:itmd_full_gA}
\end{align}
with $H_1^h = [z^2+(1-z)^2](M_1^{ns} - M_1^f) + m_Q^2(M_0^{ns} - M_0^f)$ constructed directly from the closed-form $t$-integral projections of ref.~\cite{Altinoluk:2021ygv} (with exact Taylor-continuous TMD limit as $\bk_T \to 0$). In the massless limit $m_Q \to 0$ or the high-energy-factorization limit $k_g \gg \Qs$, the $H_1^h$ term vanishes identically; in the back-to-back limit $\bk_T \to 0$, eq.~\eqref{eq:itmd_full_gA} recovers the exact unpolarized and $\cos(2\varphi)$ linearly polarized massive TMD cross section. Inside the two-dimensional transverse-momentum convolution over the target gluon momentum $\bk_g$ and soft recoil $\bq_{\mathrm{rad}} = \bk_1 + \bk_2 - \bk_g$, we evaluate eqs.~\eqref{eq:itmd_amps}--\eqref{eq:itmd_full_gA} at the momentum-conserving hard kinematics
\begin{equation}
  \tilde{\bk}_1 = \bP_T + z\bk_g, \qquad \tilde{\bk}_2 = -\bP_T + (1-z)\bk_g,
  \label{eq:kg_itmd_kin}
\end{equation}
which satisfies $\tilde{\bk}_1 + \tilde{\bk}_2 = \bk_g$ and $(1-z)\tilde{\bk}_1 - z\tilde{\bk}_2 = \bP_T$ identically.

To keep the $3\times 3$ soft-matrix contraction consistent with the off-shell kinematics~\eqref{eq:kg_itmd_kin} across the full azimuthal range, we project $\mathcal{A}_1 (t^b t^a) + \mathcal{A}_2 (t^a t^b)$ directly onto the orthonormal basis~\eqref{eq:color_basis}. Separating the symmetric and antisymmetric color combinations $\mathcal{A}_S = \mathcal{A}_1 + \mathcal{A}_2$ and $\mathcal{A}_A = \mathcal{A}_2 - \mathcal{A}_1$, the spin-summed $3\times 3$ hard color density matrix takes the exact form
\begin{equation}
  \mathbf{H}(\tilde{\bk}_1,\tilde{\bk}_2) = |\mathcal{A}_S|^2\,\bm{e}_S\bm{e}_S^T + |\mathcal{A}_A|^2\,\bm{e}_A\bm{e}_A^T + \big(|\mathcal{A}_2|^2 - |\mathcal{A}_1|^2\big)\big(\bm{e}_S\bm{e}_A^T + \bm{e}_A\bm{e}_S^T\big),
  \label{eq:H_offshell}
\end{equation}
where
\begin{equation}
  \bm{e}_S = \left(\frac{1}{2}\sqrt{\frac{\Nc^2-1}{\Nc}},\,
    \frac{1}{2}\sqrt{\frac{(\Nc^2-1)(\Nc^2-4)}{2\Nc}},\, 0\right)^{\!T},
  \qquad
  \bm{e}_A = \left(0,\, 0,\, \frac{1}{2}\sqrt{\frac{\Nc(\Nc^2-1)}{2}}\right)^{\!T}.
\end{equation}
Contracting $\mathbf{H}$ with $\mathbf{U}^\dagger\mathbf{U}$ yields the kinematics-dependent trace-normalized soft evolution factor
\begin{equation}
  R_{\mathrm{full}}(\tilde{\bk}_1,\tilde{\bk}_2; b) = \frac{\Tr\!\left[\mathbf{H}(\tilde{\bk}_1,\tilde{\bk}_2)\,\mathbf{U}^\dagger(\mu_b,\mu)\mathbf{U}(\mu_b,\mu)\right]}{\Tr\!\left[\mathbf{H}(\tilde{\bk}_1,\tilde{\bk}_2)\right]},
  \label{eq:Rfull_offshell}
\end{equation}
together with its exact off-shell planar large-$\Nc$ limit
\begin{equation}
  R_{\mathrm{planar}}(\tilde{\bk}_1,\tilde{\bk}_2; b) = \frac{|\mathcal{A}_1|^2\,e^{-4\Nc\ln(z)\,I_{\alpha_s}} + |\mathcal{A}_2|^2\,e^{-4\Nc\ln(1-z)\,I_{\alpha_s}}}{|\mathcal{A}_1|^2 + |\mathcal{A}_2|^2}.
  \label{eq:Rplanar_offshell}
\end{equation}
As $\bk_g \to 0$, $\mathcal{A}_1 \to (1-z)\mathcal{A}_0$ and $\mathcal{A}_2 \to z\mathcal{A}_0$, so eqs.~\eqref{eq:Rfull_offshell}--\eqref{eq:Rplanar_offshell} reduce identically to eqs.~\eqref{eq:Rfull_def} and~\eqref{eq:Rplanar_exact}, while away from the back-to-back limit (for instance at $z = 1/2$, $\bP_T = 0$, where $|\mathcal{M}_0(1/2)\rangle$ has zero $|8_A\rangle$ projection) eq.~\eqref{eq:H_offshell} retains the exact off-shell $|8_A\rangle$ amplitude and dilute color-trace normalization $\Tr(\mathbf{H})$. We note that eq.~\eqref{eq:Rfull_offshell} applies the trace-normalized soft factor uniformly to the dense target correlator~\eqref{eq:itmd_full_gA}; a fully color-coupled evolution of multi-Wilson-line target correlators with the soft matrix in the deeply saturated regime ($k_g \lesssim \Qs$) remains an open NLO CGC+Sudakov problem beyond the scope of current factorization theorems.

For forward $\DD$ ($m_c=1.28\gev$, $M_D=1.865\gev$) and $\BB$ ($m_b=4.50\gev$, $M_B=5.280\gev$) production at LHCb ($\sqrt{s_{NN}} = 8.16\tev$, $2.7 < y_{1,2} < 3.7$), we adopt the finite-mass collinear fragmentation kinematics, CT18NNLO collinear gluon PDFs and running $\alpha_s$~\cite{Hou:2019efy}, KKKS08 $c\to D^0$ fragmentation functions~\cite{Kneesch:2007ey}, DGLAP-evolved KKSS $b\to B^0$ fragmentation functions~\cite{Kniehl:2008zza}, and the nonperturbative CSS Sudakov exponent~\cite{Sun:2014dqm,Salazar:2026ssz}
\begin{equation}
  S_{\mathrm{NP}}(b; \mu, \CNP) = \CNP\left(\frac{\CA}{\CF} + 2\right)\left[\frac{g_1}{2}b^2 + \frac{g_2}{8}\ln\!\left(\frac{\mu^2}{Q_0^2}\right)\ln\!\left(1 + \frac{b^2}{b_{\max}^2}\right)\right],
  \label{eq:SNP}
\end{equation}
with $g_1=0.212\gev^2$, $g_2=0.84$, $Q_0^2=2.4\gev^2$, $b_{\max}=1.5\gev^{-1}$, baseline $\CNP = 2.3$, and a five-point uncertainty scan $\CNP \in [1.5, 3.5]$. To isolate target-correlator and soft-matrix finite-$\Nc$ dynamics cleanly, both the nonperturbative factor $S_{\mathrm{NP}}$ and the quark Casimir $\CF = 4/3$ inside the color-diagonal perturbative Sudakov $S_P$ are held fixed across schemes; we have verified explicitly in our archived diagnostics that replacing $\CF \to \Nc/2 = 3/2$ inside $S_P$ and $S_{\mathrm{NP}}$ shifts the finite-$\Nc$ broadening difference by only $0.11\,$pp in $\DD\,[2,4]\gev$ ($+1.19 \to +1.08\,$pp) and $0.02\,$pp in $\BB\,[4,6]\gev$ ($-0.03 \to -0.05\,$pp). Moreover, in combining $\mathbf{U}(\mu_b,\mu)$ with rcBK-evolved target TMDs, there is no double counting between wide-angle soft resummation and small-$x$ rapidity evolution: following the standard small-$x$ Sudakov factorization~\cite{Mueller:2013wwa,Caucal:2022ulg,Caucal:2023fsf,Caucal:2024bae,Caucal:2025mth,Caucal:2025dih,Balitsky:2026lop}, the one-loop matrix $\MS$ resums soft gluons with $k_\perp \ll P_T$ in the central partonic rapidity window $|y| \lesssim \ln(\sqrt{\hat{s}}/k_\perp)$ around the $Q\bar{Q}$ system, whereas rcBK evolution resums strongly ordered longitudinal $\ln(1/x_A)$ logarithms below the target shockwave rapidity cutoff.

In the Sudakov kernel we set $\mu = \sqrt{x_p x_A s_{NN}}$ and $P_{\mathrm{eff}} = (1-z)k_1 + z k_2$ (which equal $M_{Q\bar{Q}}$ and $|\bP_T|$ in the back-to-back limit $q_T = 0$), using the smooth evolution scale $\mu_b(b, \mu) = c_0 / \sqrt{b_*^2 + (c_0/\mu)^2}$ to eliminate small-$b$ derivative corners in the $J_0$ Hankel transform. Perturbative Sudakov resummation is matched onto the large-angle unresummed regime through
\begin{equation}
  g_{\mathrm{match}}(\bk_g, \bq_{\mathrm{rad}}) = \frac{M_{\mathrm{inv}}^2}{M_{\mathrm{inv}}^2 + c_{\mathrm{match}}^2 \bk_g^2}\,\exp\!\left[-\left(\frac{q_{\mathrm{rad}}}{0.55\,\mu}\right)^4\right],
  \label{eq:gmatch}
\end{equation}
with $M_{\mathrm{inv}}^2(\delta) = (\bP_T^2+m_Q^2)/[z(1-z)]$ and default $c_{\mathrm{match}}=1$ (varied over $[0.5, 2.0]$ across all six bins). The fourth-power CSS domain factor $\exp[-(q_{\mathrm{rad}}/0.55\mu)^4]$ preserves all logarithmic and $\mathcal{O}(q_{\mathrm{rad}}^2/\mu^2)$ terms in the resummation region $q_{\mathrm{rad}} \ll M_T \approx \mu/2$, while smoothly shutting off the unphysical large-$q_{\mathrm{rad}} \gtrsim \mu$ Hankel ringing tail of the $b$-space integral. As a result, the exact unclipped recoil convolution remains strictly positive across all six LHCb $p_T$ bins over the entire azimuthal range $\Delta\Phi \in [0,\pi]$, and positivity clipping alters the broadening and away-side nuclear ratios by at most $0.03\text{--}0.04\,$pp in both $\DD\,[2,4]\gev$ and $\DD\,[6,10]\gev$.

To separate leading-$\Nc$ and finite-$\Nc$ contributions with a single unified reference, we evaluate six schemes on common scrambled quasi-Monte Carlo phase-space samples (canceling statistical noise in scheme differences): (i) \emph{Large-$\Nc$ SSZ-type} (large-$\Nc$ target TMDs with color-diagonal Sudakov $R=1$, representing the color-diagonal baseline of ref.~\cite{Salazar:2026ssz} within our unified ITMD setup); (ii) \emph{Large-$\Nc$ Planar} (large-$\Nc$ target TMDs with off-shell planar soft evolution $R_{\mathrm{planar}}$); (iii) \emph{TMD-only Finite-$\Nc$} (finite-$\Nc$ Gaussian TMDs with $R=1$); (iv) \emph{Soft-only Finite-$\Nc$} (large-$\Nc$ target TMDs with the off-shell $3\times 3$ soft matrix $R_{\mathrm{full}}$); (v) \emph{Full Finite-$\Nc$} (finite-$\Nc$ Gaussian TMDs combined with $R_{\mathrm{full}}$); and (vi) \emph{Finite-$\Nc$ TMD + Planar Soft} (finite-$\Nc$ Gaussian TMDs combined with $R_{\mathrm{planar}}$). Comparing schemes (iv), (v), and (vi) against the single Large-$\Nc$ Planar baseline (ii) provides an exact unified-reference decomposition of the finite-$\Nc$ difference (which starts at $\mathcal{O}(1/\Nc^2)$ and retains all higher-order $1/\Nc^2$ terms from $\Nc = 3$), whereas Planar minus SSZ-type isolates the leading-$\Nc$ planar soft logarithm.

For nuclear minimum-bias ($\mathrm{MB}$) targets in both $^{208}\mathrm{Pb}$ ($R_{\mathrm{Pb}}=6.62\fm, a_{\mathrm{Pb}}=0.546\fm$) and $^{197}\mathrm{Au}$ ($R_{\mathrm{Au}}=6.38\fm, a_{\mathrm{Au}}=0.535\fm$), we normalize the Woods--Saxon thickness $T_A(B)$ over the full radial domain $B \in [0, 16]\fm$ ($\int \dd^2\bm{B}\,T_A(B) = A$ to better than $10^{-5}$), map the AAMQS impact-parameter-dependent $^{208}\mathrm{Pb}$ dipole family onto $T_A(B)$ with optical-Glauber continuation $S_A = S_{\mathrm{Pb}}(0)^{T_A(B)/T_{\mathrm{Pb}}(0)}$ for the $+1.76\%$ higher central thickness of $^{197}\mathrm{Au}$ ($T_{\mathrm{Au}}(0) = 2.161\fm^{-2}$ vs.\ $T_{\mathrm{Pb}}(0) = 2.124\fm^{-2}$) and continuous single-nucleon matching $N_p$ in the peripheral halo $T_A(B) < T_{\mathrm{Pb}}(6.31\fm)$ (carrying $10.8\%$ of $^{208}\mathrm{Pb}$ and $10.6\%$ of $^{197}\mathrm{Au}$ binary-collision weight, where the AAMQS edge dipole already matches $N_p$ to $1\%$). Crucially, rather than inserting a $B$-averaged dipole into the nonlinear Gaussian TMD functionals ($\mathcal{K}[\langle N_B\rangle]$), we evaluate the exact coordinate-space kernels per impact-parameter stratum and perform true binary-collision averaging $\langle\mathcal{K}[N_B]\rangle_{\mathrm{MB}} = \sum_m w_m \mathcal{K}[N_m]$ prior to the linear Hankel transforms. This yields effective saturation scales $\Qs^2(p) = 0.56\gev^2$, $\Qs^2(\mathrm{Pb}_{\mathrm{MB}}) = 1.23\gev^2$, and $\Qs^2(\mathrm{Pb}_{\mathrm{cen}}) = 1.81\gev^2$ at $Y = 4.6$ (LHCb), and $\Qs^2(p) = 0.44\gev^2$, $\Qs^2(\mathrm{Au}_{\mathrm{MB}}) = 1.01\gev^2$, and $\Qs^2(\mathrm{Au}_{\mathrm{cen}}) = 1.51\gev^2$ at $Y = 3.6$ (EIC). Comparing $\langle\mathcal{K}[N_B]\rangle_{\mathrm{MB}}$ against $\mathcal{K}[\langle N_B\rangle]$ quantifies the Jensen averaging bias: in $\DD\,[2,4]\gev$ MB $\pPb$, the broadening moves slightly from $13.15\%$ to $13.08\%$ (and the finite-$\Nc$ shift from $+0.90$ to $+0.79\,$pp), while in EIC MB $\eAu$ ($c\bar{c}$, $P_T = 3\gev, Q^2 = 2\gev^2$) per-$B$ averaging lowers the MB broadening from $9.09\%$ to $8.13\%$ while keeping the finite-$\Nc$ shift stable ($+0.47$ vs.\ $+0.51\,$pp).

For EIC $\gamma^*_{T,L}(Q^2) + A \to Q(\bk_1) + \bar{Q}(\bk_2) + X$ leptoproduction at $Y = 3.6$, we compute genuine opening-angle distributions $\dd\sigma_{\mathrm{EIC}}/\dd\Delta\Phi$ by integrating the full DIS massive ITMD cross section~\cite{Altinoluk:2021ygv} over a finite quark transverse-momentum window $k_1, k_2 \in [P_T - \Delta_p, P_T + \Delta_p]$ ($\Delta_p = 0.5\gev$, avoiding the accidental $k_1 = k_2$ vanishing of $H_L^{ns}$) with exact phase-space Jacobian $k_1 k_2\,\dd k_1\,\dd k_2$:
\begin{align}
  \frac{\dd\sigma_{\mathrm{EIC}}}{\dd\Delta\Phi} &\propto \int_{P_T-\Delta_p}^{P_T+\Delta_p} \!\! k_1\,\dd k_1 \int_{P_T-\Delta_p}^{P_T+\Delta_p} \!\! k_2\,\dd k_2 \Big[
    H_{\mathrm{DIS}}^{ns}(\bk_1,\bk_2)\,\widetilde{\mathcal{F}}_{\mathrm{WW}}(q_T) \nonumber\\
  &\qquad\qquad\qquad\qquad\qquad\qquad
    + H_{\mathrm{DIS}}^{h}(\bk_1,\bk_2)\Big(\widetilde{\mathcal{H}}_{\mathrm{WW}}(q_T) - \widetilde{\mathcal{F}}_{\mathrm{WW}}(q_T)\Big)
  \Big], \label{eq:eic_1d}
\end{align}
where $q_T = |\bk_1 + \bk_2|$, $H_{\mathrm{DIS}}^{ns,h} = H_T^{ns,h} + f_L(y)\,H_L^{ns,h}$ includes both transverse and longitudinal virtual-photon hard factors weighted by the explicit lepton inelasticity factor $f_L(y) = 2(1-y)/[1+(1-y)^2]$ (evaluated at baseline $y = 0.25$ and checked over $y \in [0.10, 0.50]$), and $\widetilde{\mathcal{F}}_{\mathrm{WW}}(q_T), \widetilde{\mathcal{H}}_{\mathrm{WW}}(q_T)$ are the Sudakov-resummed $J_0$ and $-J_2$ Hankel transforms of $\langle\mathcal{K}_{\mathrm{WW}}\rangle, \langle\mathcal{H}_{\mathrm{WW}}\rangle \times e^{-S_P^{\gamma^* A}-S_{\mathrm{NP}}^{\gamma^* A}} R_{\mathrm{EIC}}(b)$, matched smoothly via $\exp[-(q_T/0.55\mu)^4]$ onto the unresummed WW TMDs at large $q_T$. In the back-to-back regime $q_T \le P_T$, the elliptic azimuthal anisotropy $v_2(q_T) \equiv \langle\cos 2\phi_{Pq}\rangle$ is given by
\begin{equation}
  v_2(q_T) = \frac{1}{2}\frac{H_{\mathrm{lin}}^{\gamma^* A}(y)}{H_{\mathrm{unpol}}^{\gamma^* A}(y)}\,\frac{-\int \frac{b\,\dd b}{2\pi}\,J_2(q_T b)\,\langle\mathcal{H}_{\mathrm{WW}}(Y, b)\rangle\,e^{-S_P^{\gamma^* A}-S_{\mathrm{NP}}^{\gamma^* A}}\,R_{\mathrm{EIC}}(b)}{\int \frac{b\,\dd b}{2\pi}\,J_0(q_T b)\,\langle\mathcal{K}_{\mathrm{WW}}(Y, b)\rangle\,e^{-S_P^{\gamma^* A}-S_{\mathrm{NP}}^{\gamma^* A}}\,R_{\mathrm{EIC}}(b)}, \label{eq:eic_v2}
\end{equation}
where $H_{\mathrm{lin}}^{\gamma^* A}/H_{\mathrm{unpol}}^{\gamma^* A} = [2P_T^2(m_Q^2 - c_z\epsilon^2) + 8 f_L(y) z^2(1-z)^2 Q^2 P_T^2]/[c_z(P_T^4+\epsilon^4) + 2m_Q^2 P_T^2 + 8 f_L(y) z^2(1-z)^2 Q^2 P_T^2]$, $\epsilon^2 = z(1-z)Q^2 + m_Q^2$, and $c_z = z^2+(1-z)^2$. For $\gamma^*+A$ scattering, the hard resummation scale is $\mu^2 = M_{Q\bar{Q}}^2 + Q^2 = (P_T^2+\epsilon^2)/[z(1-z)]$; the single initial-state gluon halves the adjoint cusp and nonperturbative color weights to $\CA + 2\CF$ and $\CA/(2\CF)+2$, removes the incoming collinear-gluon $-2\beta_0$ term from eq.~\eqref{eq:SP_def}, and uses $\CNP = 1.0$. We restrict eq.~\eqref{eq:eic_v2} strictly to the valid back-to-back window $q_T \le \min(P_T, 3\gev)$.

\section{Phenomenological results for LHCb and the EIC}
\label{sec:results}

In this section we evaluate the phenomenological consequences of our framework for forward $\pPb$ collisions at LHCb and $\eAu$ leptoproduction at the EIC.

\subsection{Full-angle LHCb azimuthal distributions and nuclear broadening}
\label{sec:lhcb_results}

\begin{figure}[tbp]
  \centering
  \includegraphics[width=0.96\textwidth]{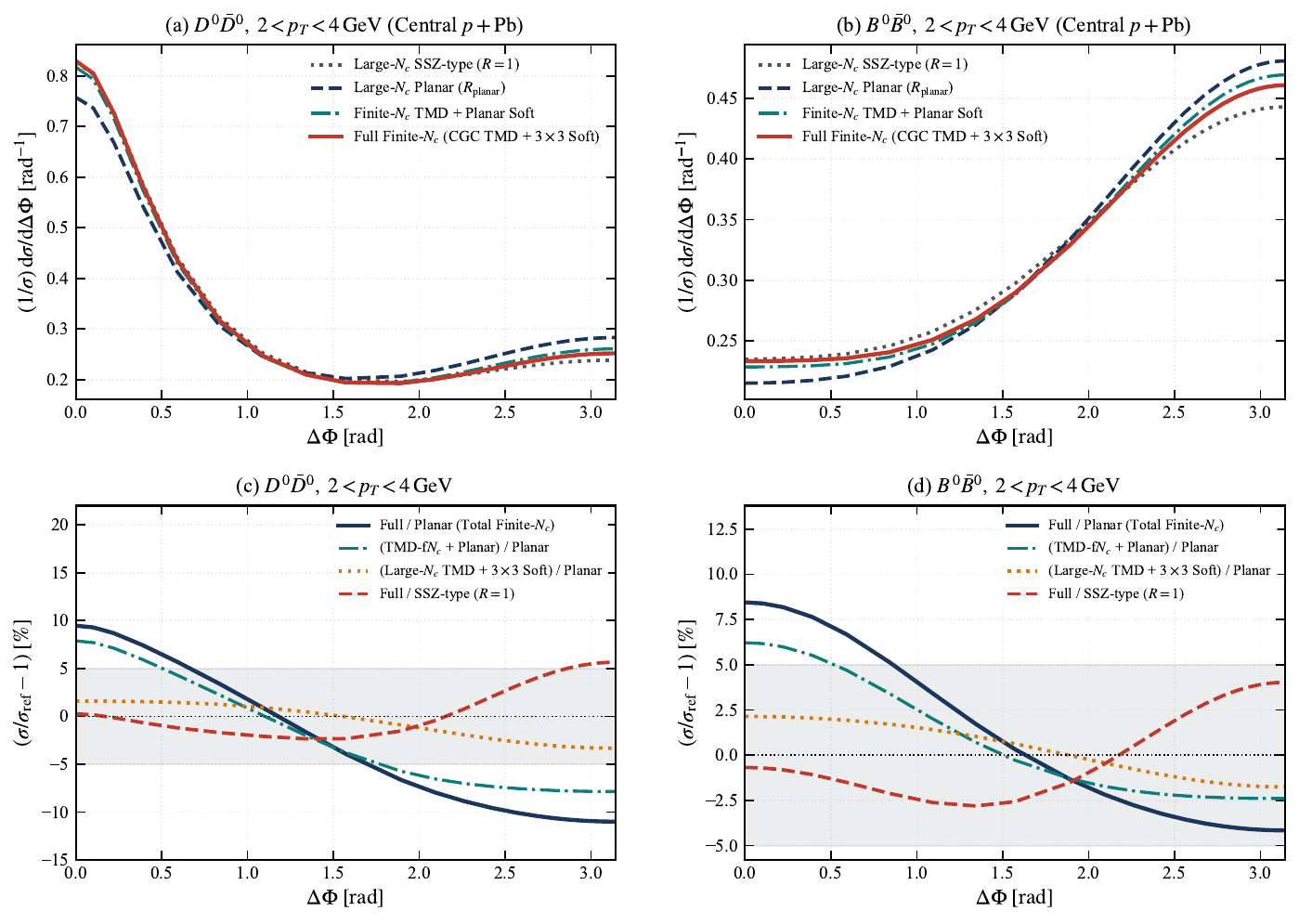}
  \caption{Full-angle normalized azimuthal distributions $(1/\sigma)\,\dd\sigma/\dd\Delta\Phi$ (panels a, b) and relative shifts $\sigma/\sigma_{\mathrm{ref}}-1$ (panels c, d) for central $\pPb$ $\DD$ (left) and $\BB$ (right) in $2 < p_T < 4\gev$ at $\sqrt{s_{NN}}=8.16\tev$, $2.7 < y_{1,2} < 3.7$, and $\CNP = 2.3$. Panels (c, d) decompose the finite-$\Nc$ shift against the unified Large-$\Nc$ Planar baseline into total (Full/Planar), target-TMD ($[\text{TMD-f}\Nc + \text{Planar}]/\text{Planar}$), and $3\times 3$ soft-matrix ($[\text{Large-}\Nc\text{ TMD} + 3\times 3\text{ Soft}]/\text{Planar}$) components, alongside the total shift relative to the SSZ-type baseline (Full/SSZ-type), with the gray shaded band marking $\pm 5\%$.}
  \label{fig:dphi}
\end{figure}

We first examine individual azimuthal distributions in central $\pPb$ collisions. Figure~\ref{fig:dphi} shows the normalized distributions for $\DD$ and $\BB$ pairs in $2 < p_T < 4\gev$, together with the unified-reference scheme-by-scheme relative shifts in the lower panels. With the full massive ITMD hard factors~\eqref{eq:itmd_full_gA}--\eqref{eq:kg_itmd_kin}, the charm spectrum exhibits both a collinear gluon-splitting peak near $\Delta\Phi = 0$ and a back-to-back peak near $\Delta\Phi = \pi$, whereas the larger bottom mass suppresses collinear splitting and leaves a distribution dominated by the away-side peak. Relative to the single Large-$\Nc$ Planar baseline, the finite-$\Nc$ difference (Full/Planar) suppresses the away-side peak at $\Delta\Phi = \pi$ by $11.0\%$ in $\DD$ ($4.2\%$ in $\BB$) and enhances the near-side region by $9.5\%$ in $\DD$ ($8.4\%$ in $\BB$). Roughly three-quarters of this redistribution comes from the finite-$\Nc$ target gluon TMDs (the $+1/\Nc^2$ WW quadrupole tail and near-side destructive interference $K_{gg}^{(2)} < 0$, contributing up to $7.9\%$ in $\DD$ and $6.2\%$ in $\BB$ against Planar), with the remaining $2\%\text{--}3.3\%$ supplied by stronger Sudakov suppression in the color-singlet channel of the off-shell $3\times 3$ soft matrix.

Comparing Full Finite-$\Nc$ directly with the color-diagonal SSZ-type baseline ($R = 1$) exposes the compensating effect of the omitted leading-$\Nc$ planar soft logarithm. Because $R_{\mathrm{planar}}(b) > 1$ sharpens and raises the away-side peak in the opposite direction to the finite-$\Nc$ corrections, the planar enhancement partially outweighs the finite-$\Nc$ suppression: the total Full/SSZ-type ratio (dashed red curves in panels~c and~d) raises the away-side peak by $5.6\%$ in low-$p_T$ $\DD$ ($4.0\%$ in $\BB$) and lowers the intermediate/near-side region by $2\%\text{--}3\%$. Omitting the leading-$\Nc$ planar soft factor, or including it without compensating finite-$\Nc$ TMD and singlet corrections, therefore modifies the individual $\pPb$ azimuthal distribution at the $5\%\text{--}11\%$ level (and up to $14\%\text{--}21\%$ in the higher-$p_T$ bins near $\Delta\Phi \to 0$).

\begin{figure}[tbp]
  \centering
  \includegraphics[width=0.98\textwidth]{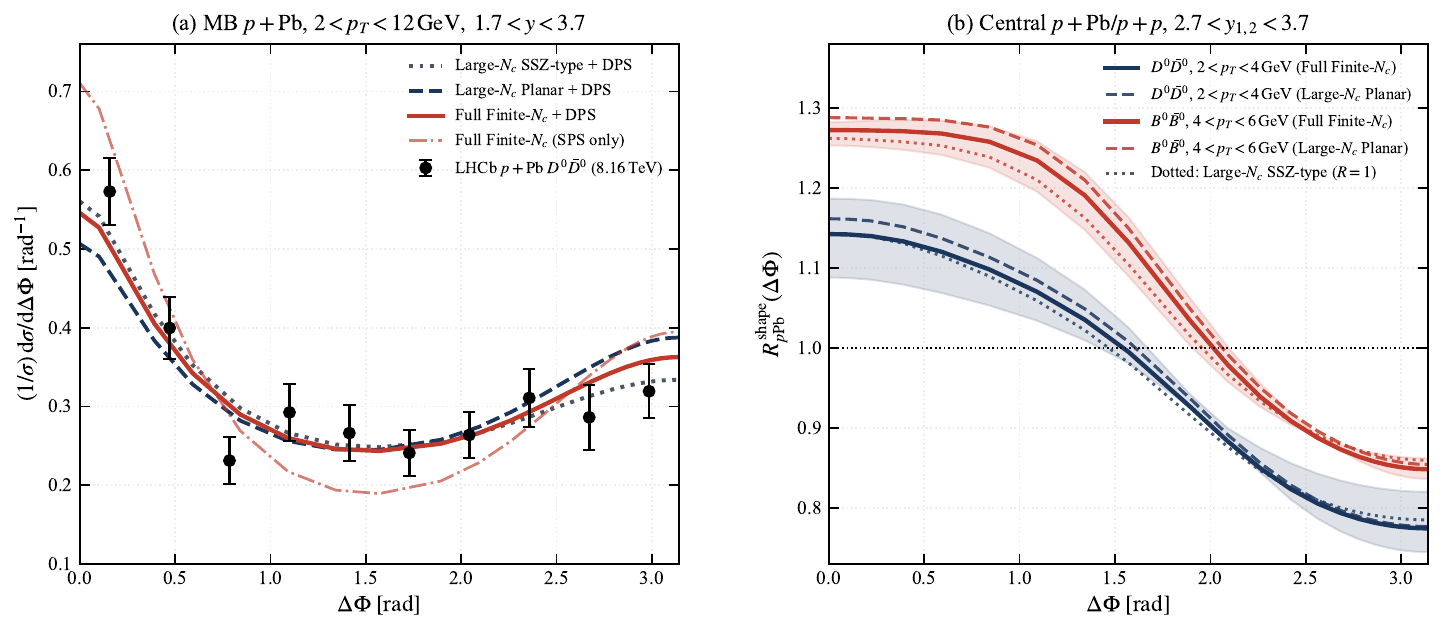}
  \caption{(a) Comparison of minimum-bias $\pPb$ $\DD$ azimuthal correlations at $\sqrt{s_{NN}} = 8.16\tev$ ($2 < p_T < 12\gev$, $1.7 < y < 3.7$) against published LHCb data~\cite{LHCb:2020jse}, including a flat Double Parton Scattering background ($f_{\mathrm{DPS}} = 0.42$) alongside the pure single-parton-scattering Full Finite-$\Nc$ curve. (b) Predictions for the LHCb Run~3--4 normalized nuclear azimuthal ratio $R_{\pPb}^{\mathrm{shape}}(\Delta\Phi)$ in central $\pPb$ relative to $\pp$ for forward $\DD$ ($2<p_T<4\gev$) and $\BB$ ($4<p_T<6\gev$), with shaded bands spanning $\CNP\in[1.5,3.5]$ for Full Finite-$\Nc$, compared with Large-$\Nc$ Planar (dashed) and SSZ-type (dotted).}
  \label{fig:lhcb_data}
\end{figure}

\begin{table}[tbp]
  \centering
  \footnotesize
  \setlength{\tabcolsep}{4.5pt}
  \renewcommand{\arraystretch}{1.24}
  \begin{tabular}{@{}lcccccc@{}}
    \toprule
    Channel ($p_T\,[\mathrm{GeV}]$) & SSZ-type ($R=1$) & Planar ($R_{\mathrm{pl}}$) & Full Finite-$\Nc$ & $\Delta_{\mathrm{planar}}$ & $\Delta_{\mathrm{fin-}\Nc}$ & $\Delta_{\mathrm{tot}}$ \\
    & $[\%]$ & $[\%]$ & $[\%]$ & $[\mathrm{pp}]$ & $[\mathrm{pp}]$ & $[\mathrm{pp}]$ \\
    \midrule
    $\DD\,[2,4]$  & $22.0^{+11.5}_{-1.2}$ & $19.2^{+3.4}_{-0.2}$ & $20.4^{+5.7}_{-0.9}$ & $-2.8^{+1.2}_{-8.1}$ & $+1.2^{+2.3}_{-0.9}$ & $-1.6^{+0.3}_{-5.8}$ \\
    $\DD\,[4,6]$  & $11.0^{+1.1}_{-0.9}$  & $11.1^{+1.2}_{-1.0}$ & $11.3^{+1.0}_{-0.8}$ & $+0.1^{+0.0}_{-0.1}$ & $+0.2^{+0.2}_{-0.2}$ & $+0.3^{+0.1}_{-0.1}$ \\
    $\DD\,[6,10]$ & $8.2^{+0.8}_{-0.8}$   & $8.6^{+1.0}_{-1.0}$  & $8.6^{+0.9}_{-0.9}$  & $+0.3^{+0.1}_{-0.1}$ & $+0.1^{+0.1}_{-0.1}$ & $+0.4^{+0.1}_{-0.0}$ \\
    \midrule
    $\BB\,[2,4]$  & $12.2^{+1.9}_{-1.0}$  & $12.2^{+1.9}_{-1.3}$ & $12.4^{+1.5}_{-0.9}$ & $-0.0^{+0.0}_{-0.3}$ & $+0.2^{+0.4}_{-0.4}$ & $+0.2^{+0.2}_{-0.5}$ \\
    $\BB\,[4,6]$  & $9.7^{+1.5}_{-1.2}$   & $10.1^{+1.5}_{-1.3}$ & $10.0^{+1.3}_{-1.1}$ & $+0.3^{+0.0}_{-0.0}$ & $-0.0^{+0.2}_{-0.2}$ & $+0.3^{+0.1}_{-0.2}$ \\
    $\BB\,[6,10]$ & $7.9^{+0.9}_{-0.9}$   & $8.2^{+1.0}_{-0.9}$  & $8.2^{+0.9}_{-0.9}$  & $+0.3^{+0.1}_{-0.0}$ & $-0.0^{+0.1}_{-0.1}$ & $+0.3^{+0.0}_{-0.0}$ \\
    \bottomrule
  \end{tabular}
  \caption{Central $\pPb$ versus $\pp$ nuclear broadening $(\sigma_{\pPb}/\sigma_{\pp} - 1)$ at $\sqrt{s_{NN}}=8.16\tev$, $2.7 < y_{1,2} < 3.7$ for baseline $\CNP = 2.3$, with asymmetric superscript/subscript envelopes spanning the five-point scan $\CNP \in [1.5, 3.5]$. The right three columns give the paired scheme differences (with their own paired five-point $\CNP$ envelopes) for the leading-$\Nc$ planar soft-logarithm shift $\Delta_{\mathrm{planar}} = \text{Planar}-\text{SSZ-type}$, the finite-$\Nc$ difference $\Delta_{\mathrm{fin-}\Nc} = \text{Full}-\text{Planar}$, and the total shift $\Delta_{\mathrm{tot}} = \text{Full}-\text{SSZ-type}$ in percentage points (pp, computed from unrounded widths; paired Jackknife statistical errors are $\le 0.04\,$pp on $\Delta_{\mathrm{fin-}\Nc}$ and $\le 0.12\,$pp on $\Delta_{\mathrm{tot}}$). Matching-scale variations $c_{\mathrm{match}} \in [0.5, 2.0]$ across all six bins preserve all qualitative trends.}
  \label{tab:lhcb_master}
\end{table}

In figure~\ref{fig:lhcb_data}(a), we compare our per-impact-parameter averaged minimum-bias $\pPb$ calculation ($2 < p_T < 12\gev$, $1.7 < y < 3.7$, $\sqrt{s_{NN}} = 8.16\tev$) with the published 10-bin LHCb $\DD$ measurement~\cite{LHCb:2020jse}. In nuclear collisions, uncorrelated Double Parton Scattering (DPS) is enhanced relative to single-parton scattering (SPS)~\cite{Helenius:2019uge} and adds an azimuthally flat background, as seen in LHCb's like-sign $D^0 D^0$ correlations~\cite{LHCb:2020jse}. Integrating each theory curve over the exact experimental bin intervals $[k\pi/10, (k+1)\pi/10]$ and combining with a common flat DPS fraction ($f_{\mathrm{DPS}} = 0.42$), our full ITMD calculation reproduces both the near-side gluon-splitting peak and the away-side peak. Under diagonal statistical uncertainties, Large-$\Nc$ Planar overshoots the away-side peak and undershoots the near side ($\chi^2/\mathrm{dof} = 2.01$, or $1.61$ when adding an illustrative $5\%$ uncorrelated systematic uncertainty), whereas restoring finite-$\Nc$ target TMDs and the $3\times 3$ soft matrix reduces this substantially to $\chi^2/\mathrm{dof} = 1.37$ ($1.13$ with $5\%$ systematic), comparable to the color-diagonal SSZ-type baseline ($\chi^2/\mathrm{dof} = 1.17$, or $0.98$ with $5\%$ systematic); optimizing $f_{\mathrm{DPS}}$ freely yields $f_{\mathrm{DPS}}^{\mathrm{opt}} \simeq 0.33\text{--}0.34$ and the same ordering ($1.22$, $1.45$, and $2.17$ for SSZ-type, Full Finite-$\Nc$, and Planar). Given the flat-DPS approximation and the precision of the 10-bin Run~2 measurement, current $\pPb$ data alone do not statistically discriminate between Full Finite-$\Nc$ and the $R = 1$ SSZ-type baseline, though they clearly disfavor adding $R_{\mathrm{planar}}$ without compensating finite-$\Nc$ corrections.

For future LHCb Run~3--4 saturation measurements, nuclear-to-proton ratios are far more decisive because collinear PDF, fragmentation, and normalization uncertainties largely cancel. Figure~\ref{fig:lhcb_data}(b) shows our predictions for the normalized nuclear azimuthal ratio
\begin{equation}
  R_{\pPb}^{\mathrm{shape}}(\Delta\Phi) \equiv \frac{\big[(1/\sigma)\,\dd\sigma/\dd\Delta\Phi\big]_{\pPb}}{\big[(1/\sigma)\,\dd\sigma/\dd\Delta\Phi\big]_{\pp}}
  \label{eq:RpPb_shape}
\end{equation}
in central $\pPb$ relative to $\pp$ collisions for $\DD$ ($2 < p_T < 4\gev$) and $\BB$ ($4 < p_T < 6\gev$), where DPS contamination is much smaller. Saturation suppresses the away-side peak near $\Delta\Phi = \pi$ by $15\%\text{--}23\%$ with a compensating enhancement toward $\Delta\Phi = 0$. In this nuclear ratio, Full Finite-$\Nc$, Large-$\Nc$ Planar, and SSZ-type all track closely across $\Delta\Phi$ and remain well within the $\CNP$ uncertainty band.

To quantify away-side broadening across kinematics, we extract the Gaussian width $\sigma$ on $|\Delta\Phi - \pi| \le 0.70\,\mathrm{rad}$ for all six LHCb $p_T$ and flavor bins; the central $\pPb/\pp$ broadening results and paired $\CNP$ shift envelopes are listed in table~\ref{tab:lhcb_master}. Across all $\BB$ bins and all $\DD$ bins with $p_T \ge 4\gev$, the finite-$\Nc$ difference ($\Delta_{\mathrm{fin-}\Nc} = \text{Full}-\text{Planar}$) cancels to the sub-percentage-point level: $+0.22\,$pp ($-0.20$ to $+0.65\,$pp across $\CNP$) for $\BB\,[2,4]\gev$, $+0.19\,$pp for $\DD\,[4,6]\gev$, $+0.05\,$pp for $\DD\,[6,10]\gev$, and $-0.03\,$pp / $-0.01\,$pp for $\BB\,[4,6]$ and $[6,10]\gev$ (with $\le 0.18\,$pp in minimum-bias $\pPb$). Three effects drive this cancellation: overall color prefactors drop out of the normalized $\pPb/\pp$ width ratio, low-$k_T$ enhancement from the nuclear WW quadrupole tail partially compensates both the larger adjoint Casimir ratio and stronger Sudakov suppression in the color-singlet channel, and finite-$\Nc$ corrections decouple once $p_T$ exceeds $\Qs$. Only in the lowest charm bin $\DD\,[2,4]\gev$, where the away-side peak is intrinsically wide ($\sigma \gtrsim 1.1\,\mathrm{rad}$) and sensitive to the near-side collinear tail, does $\Delta_{\mathrm{fin-}\Nc}$ reach $+1.19\,$pp ($+0.30$ to $+3.48\,$pp across $\CNP$, and $+0.79\,$pp in minimum bias). Varying the Gaussian fit window over $|\Delta\Phi - \pi| \le 0.50\text{--}0.90\,\mathrm{rad}$ changes $\Delta_{\mathrm{fin-}\Nc}$ mildly (from $+1.14$ to $+1.48\,$pp in $\DD\,[2,4]\gev$, and staying within $\pm 0.33\,$pp in all other bins), while varying $c_{\mathrm{match}} \in [0.5, 2.0]$ across all six bins preserves every qualitative trend (yielding $\Delta_{\mathrm{fin-}\Nc} = +1.15\text{ to }+1.31\,$pp in $\DD\,[2,4]\gev$ and $\le 0.08\,$pp in $[6,10]\gev$). Omitting the $H_1^h(\mathcal{H}-\mathcal{F})$ ITMD term (the ITMD$^*$ approximation) lowers the overall central broadening in $\DD\,[2,4]\gev$ ($20.39\% \to 17.58\%$) and $\BB\,[2,4]\gev$ ($12.37\% \to 9.70\%$) while leaving the finite-$\Nc$ shift remarkably stable ($+1.21\,$pp in $\DD$ and $+0.39\,$pp in $\BB$).

Table~\ref{tab:lhcb_master} also shows that in $\DD\,[2,4]\gev$, the leading-$\Nc$ planar soft logarithm produces a noticeable downward shift ($\Delta_{\mathrm{planar}} = -2.78\,$pp at $\CNP = 2.3$, expanding to $-10.8\,$pp at $\CNP = 3.5$) and stabilizes the large-$\CNP$ away-side width envelope ($+11.5\,$pp in SSZ-type reduced to $+3.4\,$pp in Planar and $+5.7\,$pp in Full Finite-$\Nc$), whereas in all other bins $|\Delta_{\mathrm{planar}}| \le 0.34\,$pp. For all $\BB$ bins and for $\DD$ at $p_T \ge 4\gev$, large-$\Nc$ broadening predictions therefore sit well within the few-percent nuclear saturation uncertainty band of ref.~\cite{Salazar:2026ssz}.

As a non-Gaussian cross-check---especially for the broad $\DD\,[2,4]\gev$ peak---we also evaluate the finite-angle bin-integrated away-side nuclear ratio
\begin{equation}
  R_{\pPb}^{\mathrm{away}} \equiv \frac{\int_{|\Delta\Phi-\pi|\le 0.5} \dd\Delta\Phi\,(1/\sigma)\,\dd\sigma_{\pPb}/\dd\Delta\Phi}{\int_{|\Delta\Phi-\pi|\le 0.5} \dd\Delta\Phi\,(1/\sigma)\,\dd\sigma_{\pp}/\dd\Delta\Phi},
\end{equation}
alongside the peak ratio $R_{\pPb}^{\mathrm{shape}}(\pi)$ and the azimuthal decorrelation $\Delta\langle\cos\delta\rangle = \langle\cos\delta\rangle_{\pPb} - \langle\cos\delta\rangle_{\pp}$. In $\DD\,[2,4]\gev$ at $\CNP = 2.3$, $R_{\pPb}^{\mathrm{away}}$ moves from $0.791$ (SSZ-type) to $0.784$ (Planar, $-0.65\,$pp) and $0.782$ (Full, a finite-$\Nc$ shift of only $-0.26\,$pp), matching the peak ratio $R_{\pPb}^{\mathrm{shape}}(\pi)$ ($0.785 \to 0.777 \to 0.774$, finite-$\Nc$ shift $-0.22\,$pp). Across the remaining five bins, the finite-$\Nc$ shift on $R_{\pPb}^{\mathrm{away}}$ remains between $-0.22\,$pp ($\BB\,[2,4]\gev$) and $-1.35\,$pp ($\DD\,[6,10]\gev$), while $\Delta\langle\cos\delta\rangle$ changes by less than $0.008$ in every bin. Both width-based and bin-integrated observables thus support the same physical picture: color-correlated soft resummation and finite-$\Nc$ corrections are important for individual $\pPb$ azimuthal shapes and for low-$p_T$ charm broadening, whereas normalized $\pPb/\pp$ ratios at $p_T \ge 4\gev$ or in bottom-pair production are robust against finite-$\Nc$ corrections.

\subsection{EIC heavy-quark correlations and elliptic azimuthal anisotropy}
\label{sec:eic_results}

\begin{figure}[tbp]
  \centering
  \includegraphics[width=\textwidth]{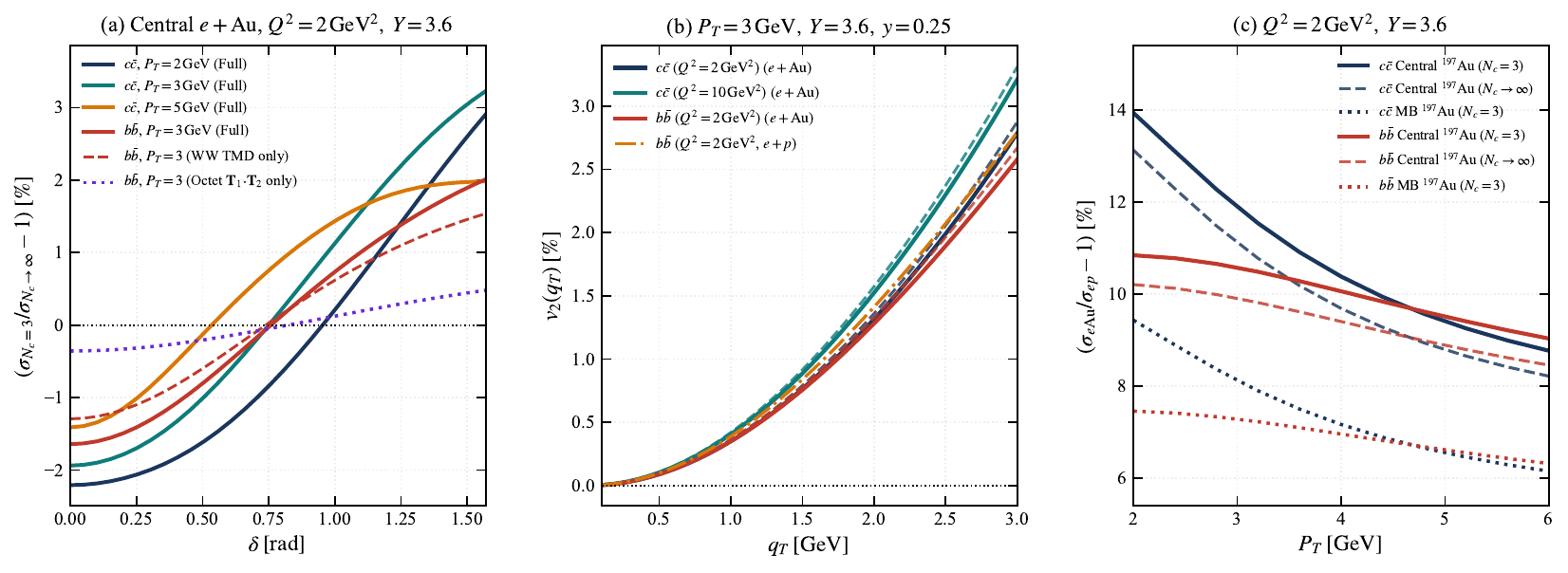}
  \caption{EIC ($\ep$ vs.\ per-impact-parameter averaged optical-Glauber $^{197}\mathrm{Au}$) finite-$\Nc$ WW quadrupole and 1D octet soft-dipole corrections at $Y=3.6$, $y = 0.25$, using full DIS massive ITMD hard factors integrated over $k_{1,2} \in [P_T - 0.5, P_T + 0.5]\gev$: (a) WW TMD versus octet $\mathbf{T}_1\cdot\mathbf{T}_2$ decomposition of the normalized azimuthal shape shift $\sigma_{\Nc=3}/\sigma_{\Nc\to\infty}-1$ in central $\eAu$ at $Q^2=2\gev^2$; (b) back-to-back elliptic azimuthal anisotropy $v_2(q_T) = \langle\cos 2\phi_{Pq}\rangle$ at $P_T = 3\gev$ over the valid back-to-back window $q_T \le P_T$, comparing $\Nc=3$ (solid curves for $\eAu$, dash-dotted for $\ep$) against $\Nc\to\infty$ (dashed curves); (c) continuous $P_T$ dependence of EIC nuclear broadening $(\sigma_{e\mathrm{Au}}/\sigma_{ep} - 1)$ for $c\bar{c}$ and $b\bar{b}$ in central and minimum-bias $^{197}\mathrm{Au}$ at $Q^2 = 2\gev^2$.}
  \label{fig:eic}
\end{figure}

We next evaluate EIC heavy-quark leptoproduction ($\gamma^*_{T,L} + A \to Q\bar{Q}$) via eqs.~\eqref{eq:eic_1d}--\eqref{eq:eic_v2} on per-impact-parameter averaged optical-Glauber $^{197}\mathrm{Au}$ dipoles at $Y = 3.6$ and inelasticity $y = 0.25$. The results are collected in figure~\ref{fig:eic}.

Unlike in hadronic $\pPb$ collisions, where low-$k_T$ enhancement from the nuclear WW quadrupole tail and stronger Sudakov suppression in the color-singlet channel act in opposite directions, the two surviving finite-$\Nc$ effects in $\gamma^* + A \to Q\bar{Q}$ add with the same sign. In the nuclear gluon distribution, $\FWW$ receives the $+12.5\%$ increase of the adjoint Casimir ratio ($\alpha = 9/4$ versus $2$) inside $W(r)$, mildly depleting low-$k_T$ gluons and broadening the away-side peak; in soft-gluon evolution, the repulsive color-octet heavy-quark dipole $-\frac{2}{\Nc}\ReP f(\beta) > 0$ in eq.~\eqref{eq:REIC_exact} slightly increases soft radiation. Even when both effects combine, panel~(a) shows that the normalized azimuthal shape shift in central $\eAu$ stays within $-2.2\%$ to $+3.2\%$ over $\delta \le \pi/2$ (and below $5.8\%$ over the full range $\Delta\Phi \in [0,\pi]$) for both $c\bar{c}$ and $b\bar{b}$ across $P_T = 2\text{--}5\gev$, with the octet $\mathbf{T}_1\cdot\mathbf{T}_2$ dipole contributing less than $0.5\%$ for $b\bar{b}$.

Panel~(b) shows the back-to-back elliptic azimuthal anisotropy $v_2(q_T)$ computed from eq.~\eqref{eq:eic_v2} with $\mathcal{H}_{\mathrm{WW}}(r) = W(r)\Dlin L(r)$ on $q_T \in [0.1, 3.0]\gev$ at $P_T = 3\gev$~\cite{Boer:2010zf,Metz:2011wb,Dominguez:2011br,Pisano:2013cya,Dumitru:2015gaa,Boer:2016fqd,Dumitru:2018kuw}. As expected, $v_2(q_T)$ grows with $q_T$ up to $2.6\%\text{--}3.2\%$ at $q_T = 3\gev$, increases with $Q^2$ as the longitudinal photon channel turns on, and is suppressed in $\eAu$ relative to $\ep$ by saturation. Going from $\Nc \to \infty$ (dashed) to $\Nc = 3$ (solid) reduces $v_2(q_T)$ by only $3\%\text{--}4\%$ relative across the full back-to-back $q_T$ range, leaving the nuclear suppression pattern unchanged.

Panel~(c) displays the continuous $P_T$ dependence of EIC nuclear broadening $(\sigma_{e\mathrm{Au}}/\sigma_{ep} - 1)$ for $c\bar{c}$ and $b\bar{b}$ in central and minimum-bias $^{197}\mathrm{Au}$. With the phase-space-integrated full DIS ITMD hard factors~\eqref{eq:eic_1d} and hard scale $\mu^2 = M_{Q\bar{Q}}^2 + Q^2$, EIC nuclear broadening decreases smoothly with $P_T$, quark mass, and virtuality $Q^2$ (for instance, from $11.90\%$ at $Q^2 = 2\gev^2$ to $11.53\%$ at $Q^2 = 10\gev^2$ for $c\bar{c}$ at $P_T = 3\gev$). Across $2 \le P_T \le 6\gev$ and all scanned $Q^2 \in [0.1, 10]\gev^2$, the finite-$\Nc$ shift remains modest and nearly uniform at $+0.62\text{ to }+0.82\,$pp in central $^{197}\mathrm{Au}$ and $+0.41\text{ to }+0.54\,$pp in minimum-bias $^{197}\mathrm{Au}$. We verified explicitly that varying the lepton inelasticity over $y \in [0.10, 0.50]$ changes the broadening shift by less than $0.01\,$pp, while testing asymmetric momentum sharing ($z = 0.25$ at $P_T = 3\gev$ for $c\bar{c}$ at $Q^2 = 2, 10\gev^2$ and $b\bar{b}$ at $Q^2 = 2\gev^2$) yields nearly identical finite-$\Nc$ broadening shifts ($+0.65\text{ to }+0.71\,$pp in central $\eAu$ and $+0.43\text{ to }+0.47\,$pp in minimum bias). By contrast, omitting the $H_{T,L}^h(\mathcal{H}_{\mathrm{WW}} - \FWW)$ ITMD term (ITMD$^*$) substantially underestimates the absolute nuclear broadening ($11.90\% \to 7.79\%$ at $Q^2 = 2\gev^2$, and $11.53\% \to 6.85\%$ at $Q^2 = 10\gev^2$), confirming the importance of the full DIS ITMD hard factor noted in ref.~\cite{Altinoluk:2021ygv}.

Comparing sections~\ref{sec:lhcb_results} and~\ref{sec:eic_results} summarizes large-$\Nc$ accuracy across the two colliders. In forward $\pPb$ collisions at LHCb, individual $\pPb$ azimuthal distributions receive $4\%\text{--}11\%$ finite-$\Nc$ redistributions in $2 < p_T < 4\gev$ (and up to $14\%\text{--}21\%$ in higher-$p_T$ near-side regions), while in normalized $\pPb/\pp$ nuclear broadening ratios finite-$\Nc$ shifts cancel down to $\le 0.22\,$pp for all $\BB$ bins and $p_T \ge 4\gev$ $\DD$ bins (and $\le 1.35\,$pp on bin-integrated away-side ratios across all bins). In EIC $\eAu$ leptoproduction, where $\mathcal{F}_{gg}^{(1,2)}$ mixing and singlet--octet soft evolution are absent, finite-$\Nc$ effects remain within $-2.2\%$ to $+3.2\%$ over $\delta \le \pi/2$ (and $\le 5.8\%$ over $\Delta\Phi \in [0,\pi]$) and stay uniformly near $+0.62\text{ to }+0.82\,$pp in central nuclear broadening across $2 \le P_T \le 6\gev$.

\section{Summary and outlook}
\label{sec:conclusions}

We have formulated and evaluated a finite-$\Nc$ framework for heavy-quark pair azimuthal correlations in forward $\pPb$ collisions at the LHC and $\eAu$ collisions at the EIC, treating multi-point Wilson-line correlators in the target and color-correlated wide-angle soft radiation in Sudakov resummation at matching color accuracy.

In the small-$x$ CGC sector, a Duhamel expansion of the exact $2\times 2$ Gaussian evolution matrix yields closed-form coordinate-space kernels at arbitrary $\Nc$ for both unpolarized ($\mathcal{K}_{\mathrm{dip}}, \mathcal{K}_{\mathrm{WW}}, \mathcal{K}_{gg}^{(1,2)}$) and linearly polarized ($\mathcal{H}_{\mathrm{dip}}, \mathcal{H}_{\mathrm{WW}}, \mathcal{H}_{gg}^{(1,2)}$) small-$x$ gluon TMDs. In both polarization sectors, the $(L')^2$ terms cancel in the sum of color-flow kernels, leaving a $+1/\Nc^2$ Weizs\"acker--Williams quadrupole tail in the effective $g+g^*\to Q\bar{Q}$ gluon TMD correlator. At $\Nc = 3$, this tail competes against the $+12.5\%$ increase of the adjoint Casimir ratio ($\CA/\CF = 9/4$ versus $2$): in the unpolarized sector it enhances low-$k_T$ gluons more strongly in Pb than in $p$, while in the linearly polarized sector it lifts the exponential saturation suppression of $\mathcal{W}_{gA}$ and nearly doubles $h_1^\perp/f_1$ in central Pb around $k_T \sim \Qs$.

For perturbative Sudakov resummation, direct evaluation of the one-loop four-parton real eikonal spectrum fixes all signs and coefficients of the $3\times 3$ soft anomalous dimension matrix $\MS(z, P_T, m_Q; \Nc)$ for $g+g^*\to Q\bar{Q}$. Evolving continuously with the velocity-dependent heavy-quark dipole $f(\beta)$, the matrix reproduces massless collinear cancellation ($\ReP f(\beta)\to 0$), threshold diagonalization with static color-singlet zero radiation ($S_P^{\mathrm{const}} + \ReP M_{11} = 0$) and static octet screening ($\ReP(M_{88}-M_{11}) = -2\CA$), and 't~Hooft $1/\Nc^2$ convergence to the leading-$\Nc$ planar factor $R_{\mathrm{planar}}(b) > 1$ induced by the scale reduction $\hat{s}\to -t_1,-u_1$. Away from $z = 1/2$, the Baker--Campbell--Hausdorff expansion shows that the non-Abelian Coulomb/Glauber imaginary phase cancels through two loops in real color states and enters first at three-loop order via the exact double commutator $-\frac{I_{\alpha_s}^3}{24}\langle\mathcal{M}_0|[\mathbf{B},[\mathbf{A},\mathbf{B}]]|\mathcal{M}_0\rangle/\langle\mathcal{M}_0|\mathcal{M}_0\rangle$, suppressing $R_{\mathrm{full}}(b)$ by $4.5\%\text{--}6.8\%$ at $z = 0.25, b = 1\gev^{-1}$. In EIC leptoproduction, replacing the incoming gluon by a virtual photon reduces soft evolution to a one-dimensional color-octet exponential $R_{\mathrm{EIC}}(b)$ containing only a weak repulsive heavy-quark dipole $-\frac{2}{\Nc}\ReP f(\beta) > 0$.

Combining both sectors with rcBK dipoles, per-impact-parameter Woods--Saxon nuclear averaging, full massive off-shell ITMD hard factors (including $H^h(\mathcal{H}-\mathcal{F})$), off-shell $3\times 3$ hard--soft color contraction, and smooth unclipped CSS recoil matching clarifies the phenomenological role of finite-$\Nc$ effects. In LHCb central $\pPb$ collisions, finite-$\Nc$ differences ($\text{Full}-\text{Planar}$) shift individual $\pPb$ azimuthal distributions by $4\%\text{--}11\%$ in $2 < p_T < 4\gev$, and restoring them substantially improves the bin-averaged comparison to published LHCb minimum-bias $\pPb$ $\DD$ data relative to Large-$\Nc$ Planar ($\chi^2/\mathrm{dof} = 1.37$ vs.\ $2.01$ at $f_{\mathrm{DPS}} = 0.42$), although current 10-bin data with a flat-DPS ansatz do not statistically distinguish Full Finite-$\Nc$ from the color-diagonal SSZ-type baseline ($\chi^2/\mathrm{dof} = 1.17$). In normalized $\pPb/\pp$ nuclear ratios such as Run~3--4 $R_{\pPb}^{\mathrm{shape}}(\Delta\Phi)$ and away-side Gaussian broadening, finite-$\Nc$ shifts cancel strongly across all $\BB$ bins ($\le 0.22\,$pp) and all $\DD$ bins with $p_T \ge 4\gev$ ($\le 0.19\,$pp), reaching $+1.19\,$pp only in the broad low-$p_T$ charm bin $\DD\,[2,4]\gev$ (where the bin-integrated away-side ratio $R_{\pPb}^{\mathrm{away}}$ still shifts by only $-0.26\,$pp). In $\DD\,[2,4]\gev$, the main departure from the color-diagonal SSZ-type baseline ($\Delta_{\mathrm{planar}} = -2.78\,$pp at $\CNP = 2.3$) is instead the omitted leading-$\Nc$ planar soft logarithm, which also curbs excessive large-$\CNP$ broadening of the low-$p_T$ charm away-side peak.

In EIC $\gamma^*_{T,L} + A \to Q\bar{Q}$ leptoproduction on per-impact-parameter averaged $^{197}\mathrm{Au}$ targets, the absence of $\mathcal{F}_{gg}^{(1,2)}$ mixing and singlet--octet evolution keeps finite-$\Nc$ corrections within $-2.2\%$ to $+3.2\%$ over $\delta \le \pi/2$ (and $\le 5.8\%$ over $\Delta\Phi \in [0,\pi]$), shifts the elliptic azimuthal anisotropy $v_2(q_T)$ on $q_T \le P_T$ by only $3\%\text{--}4\%$ relative, and shifts nuclear broadening uniformly by $+0.62\text{ to }+0.82\,$pp (central) and $+0.41\text{ to }+0.54\,$pp (minimum bias) across $2 \le P_T \le 6\gev$, $Q^2 \in [0.1, 10]\gev^2$, $y \in [0.10, 0.50]$, and asymmetric $z = 0.25$. Large-$\Nc$ calculations of EIC heavy-quark pair correlations are therefore quantitatively robust for small-$x$ saturation studies, provided the full DIS ITMD hard factors including $H_{T,L}^h(\mathcal{H}_{\mathrm{WW}} - \FWW)$ are retained.

Several extensions remain open for future work. Benchmarking our analytical Gaussian kernels directly against full finite-$\Nc$ Langevin JIMWLK evolution~\cite{Rummukainen:2003ns,Marquet:2016cgx,Penttala:2025tmp,Cougoulic:2026sww,Iancu:2026cgc} will quantify non-Gaussian small-$x$ corrections in the quadrupole and WW tails. Developing a fully color-coupled NLO CGC+Sudakov matching in the deeply saturated regime ($k_g \lesssim \Qs$) will extend the off-shell hard--soft contraction beyond the trace-normalized approximation used here. Since the linear polarization ratio $h_1^\perp/f_1$ of $\mathcal{W}_{gA}$ nearly doubles at $\Nc = 3$ in central Pb, a dedicated polarized resummation study is warranted to test whether a measurable $\cos(2\phi_{Pq})$ asymmetry survives nonperturbative Sudakov suppression in forward LHCb $\DD$ and $\BB$ production~\cite{Marquet:2017xwy}. Finally, combining this color-space resummation with next-to-leading-order massive CGC impact factors~\cite{Beuf:2022ndu} and two-loop three-parton massive soft correlations near threshold~\cite{Kidonakis:2009ev,Mitov:2009sv,Ferroglia:2009ep,Ferroglia:2009ii,Catani:2018mei}, together with improved experimental separation of DPS backgrounds in LHCb Run~3--4 $\BB$ and central $\pPb$ measurements, will further reduce theoretical uncertainties in heavy-quark saturation observables.

\acknowledgments
This work is supported in part by the National Natural Science Foundation of China (NSFC) under Grant No.~1234710148, and in part by the China Postdoctoral Science Foundation under Grant No.~2023M742098.

\appendix
\section{Duhamel derivation of the finite-\texorpdfstring{$N_c$}{Nc} TMD tensor kernels}
\label{app:duhamel}

Here we derive the unpolarized and linearly polarized finite-$\Nc$ Gaussian TMD kernels in eqs.~\eqref{eq:Kdip}--\eqref{eq:Hgg2} from the $2\times 2$ evolution matrix $\mathbf{M}(\bx_1,\bx_2,\bx_3,\bx_4)$ of eq.~\eqref{eq:gauss_M}. Expanding $\mathbf{M}(\bu_1, \bu_2) = \mathbf{M}_0 + u_1^i \mathbf{A}_i + u_2^j \mathbf{B}_j + u_1^i u_2^j \mathbf{C}_{ij} + \dots$ around the coincident configuration $\bx_1=\bx_4=\bb$, $\bx_2=\bx_3=\bb'$ ($\br = \bb - \bb'$), Duhamel's formula gives
\begin{align}
  \left.\frac{\partial^2 e^{\mathbf{M}}}{\partial u_1^i \partial u_2^j}\right|_{0}
  &= \int_0^1 \dd s\, e^{(1-s)\mathbf{M}_0}\,\mathbf{C}_{ij}\,e^{s\mathbf{M}_0} \nonumber\\
  &\quad + \int_0^1 \dd s \int_0^s \dd t \left[
    e^{(1-s)\mathbf{M}_0}\mathbf{A}_i e^{(s-t)\mathbf{M}_0}\mathbf{B}_j e^{t\mathbf{M}_0}
    + e^{(1-s)\mathbf{M}_0}\mathbf{B}_j e^{(s-t)\mathbf{M}_0}\mathbf{A}_i e^{t\mathbf{M}_0}
  \right].
  \label{eq:duhamel}
\end{align}

For $\mathcal{K}_{gg}^{(1),ij}(\br)$ ($\bx_1 = \bb + \bu_1$, $\bx_2 = \bb' + \bu_2$, $\bx_3 = \bb'$, $\bx_4 = \bb$), differentiating eq.~\eqref{eq:gauss_M} at $\bu_1=\bu_2=0$ yields first-derivative matrices strictly proportional to $\mathbf{M}_0$,
\begin{equation}
  \mathbf{A}_i = -\mathbf{B}_i = \frac{\Gamma'(r)\,\hat{r}_i}{2\Gamma(r)}\,\mathbf{M}_0,
  \qquad [\mathbf{A}_i, \mathbf{M}_0] = [\mathbf{B}_j, \mathbf{M}_0] = 0,
  \label{eq:commute_AB}
\end{equation}
together with the mixed second-derivative matrix
\begin{equation}
  \mathbf{C}_{ij} = \frac{2}{\CF}\,\big(\partial_i\partial_j L(\br)\big)
  \begin{pmatrix} \frac{\CF}{2} & \frac{1}{4} \\ 0 & -\frac{1}{4\Nc} \end{pmatrix}.
  \label{eq:Cij_gg1}
\end{equation}
Because $\mathbf{A}_i$ and $\mathbf{B}_j$ commute with $\mathbf{M}_0$, the double integral reduces immediately to
\begin{equation}
  \mathbf{A}_i \mathbf{B}_j e^{\mathbf{M}_0}
  = -\frac{(\alpha L')^2}{4\lambda^2}\hat{r}_i\hat{r}_j\,\mathbf{M}_0^2 e^{\mathbf{M}_0},
  \qquad
  \mathbf{M}_0^2 e^{\mathbf{M}_0} = \lambda^2 S_A \begin{pmatrix} 1 & 0 \\ -1/\Nc & 0 \end{pmatrix}.
\end{equation}
Projecting onto the double-dipole row $\bm{w}_{22}^T = (1,\, 1/\Nc)$ of eq.~\eqref{eq:S22_exact} gives the $(L')^2$ contribution
\begin{equation}
  \bm{w}_{22}^T (\mathbf{A}_i\mathbf{B}_j e^{\mathbf{M}_0})_{:,1}
  = -\frac{\alpha^2 (L')^2}{4}\left(1 - \frac{1}{\Nc^2}\right) S_A\,\hat{r}_i\hat{r}_j
  = -\frac{\Nc^2}{\Nc^2 - 1}\,S_A (L')^2\,\hat{r}_i\hat{r}_j,
  \label{eq:lp2_piece}
\end{equation}
while evaluating the single Duhamel integral using $\int_0^1 \dd s\,e^{-\lambda s} = W(r)$ yields
\begin{equation}
  \mathcal{K}_{gg}^{(1),ij}(\br) = \left[\left(1 - \frac{2}{\Nc^2}\right)S_A(r) + \frac{2}{\Nc^2}W(r)\right]\partial_i\partial_j L(\br)
  - \frac{\Nc^2}{\Nc^2 - 1}\,S_A(r)\,\big(L'(r)\big)^2\,\hat{r}_i\hat{r}_j.
  \label{eq:Kgg1_tensor}
\end{equation}
Repeating the same steps for $\mathcal{K}_{gg}^{(2),ij}$ ($\bx_1=\bb+\bu_1$, $\bx_2=\bb'$, $\bx_3=\bb'+\bu_2$, $\bx_4=\bb$, with an overall minus sign) leaves $\mathbf{A}_i$ and $\mathbf{B}_j$ unchanged---so the overall minus sign reverses the $(L')^2$ term---while the single Duhamel integral of $-\partial_{u_1^i}\partial_{u_2^j}\mathbf{M}$ projects onto $\frac{2}{\Nc^2}[W(r) - S_A(r)]\,\partial_i\partial_j L(\br)$. For the WW correlator $\mathcal{K}_{\mathrm{WW}}^{ij}$ ($\bx_1=\bb+\bu_1$, $\bx_2=\bb$, $\bx_3=\bb'$, $\bx_4=\bb'+\bu_2$, expanded around $\tilde{\mathbf{M}}_0 = \begin{psmallmatrix}0 & \lambda/\Nc \\ 0 & -\lambda\end{psmallmatrix}$ and projected onto the quadrupole row $\bm{w}_4^T = (1, \Nc)$), the first derivatives satisfy $\tilde{\mathbf{A}}_i = -\tilde{\mathbf{B}}_i \propto \tilde{\mathbf{M}}_0$ with $\tilde{\mathbf{M}}_0\bm{e}_1 = \bm{0}$, eliminating the double integral identically and leaving $\mathcal{K}_{\mathrm{WW}}^{ij}(\br) = W(r)\,\partial_i\partial_j L(\br)$. Decomposing $\partial_i\partial_j L(\br) = \frac{1}{2}\delta_{ij}\nabla_r^2 L + (\hat{r}_i\hat{r}_j - \frac{1}{2}\delta_{ij})\Dlin L$ and $\hat{r}_i\hat{r}_j = \frac{1}{2}\delta_{ij} + (\hat{r}_i\hat{r}_j - \frac{1}{2}\delta_{ij})$ immediately yields eqs.~\eqref{eq:Kdip}--\eqref{eq:Hgg2}.

\section{Four-parton one-loop eikonal integrals}
\label{app:eikonal}

In the partonic center-of-mass frame of $g(p_a) + g(p_b) \to Q(p_1) + \bar{Q}(p_2)$ with $\hat{s} = (P_T^2+m_Q^2)/[z(1-z)]$ and heavy-quark rapidities $y_1 = -y_2 = \frac{1}{2}\ln[z/(1-z)]$, the one-loop real soft gluon emission cross section at fixed transverse momentum $k_\perp \ll m_Q, P_T$ is controlled by the phase-space integrals
\begin{equation}
  I_{IJ} \equiv \int_{-Y_{\max}}^{Y_{\max}} \dd y \int_0^{2\pi} \frac{\dd\phi_k}{2\pi}\, k_\perp^2\,\frac{p_I \cdot p_J}{(p_I\cdot k)(p_J\cdot k)},
  \qquad Y_{\max} = \ln\frac{\sqrt{\hat{s}}}{k_\perp},
  \label{eq:I_IJ_def}
\end{equation}
with $k^\mu = k_\perp(\cosh y, \sinh y, \cos\phi_k, \sin\phi_k)$, entering the Sudakov exponent in $\frac{\alpha_s}{4\pi}\frac{\dd k_\perp^2}{k_\perp^2}$ units as $S_{\mathrm{eik}} = -\sum_{I,J}(\mathbf{T}_I\cdot\mathbf{T}_J)\,I_{IJ}$. Performing the azimuthal average analytically gives, for the massive diagonal terms ($I=J=1,2$),
\begin{equation}
  I_{11} = I_{22} = \int_{-\infty}^\infty \dd u\,\frac{m_Q^2 m_T\cosh u}{(m_T^2\cosh^2 u - P_T^2)^{3/2}} = \int_{-\infty}^\infty \frac{m_Q^2\,\dd\xi}{(\xi^2 + m_Q^2)^{3/2}} = 2,
\end{equation}
and for the beam--heavy-quark interference ($I=a, J=1$),
\begin{equation}
  \begin{aligned}
    I_{a1} &= \int_{-\infty}^{Y_{\max}-y_1} \dd u\,\frac{m_T e^u}{\sqrt{m_T^2\cosh^2 u - P_T^2}}
    = \ln\!\left(\frac{e^{2(Y_{\max}-y_1)}m_T^2}{m_Q^2}\right) \\
    &= \ln\frac{\hat{s}}{k_\perp^2} + \ln\frac{\hat{s}}{m_Q^2} + 2\ln(1-z).
  \end{aligned}
\end{equation}
By symmetry, $I_{b2} = I_{a1}$ and $I_{a2} = I_{b1} = \ln(\hat{s}/k_\perp^2) + \ln(\hat{s}/m_Q^2) + 2\ln z$, while $I_{ab} = 2\ln(\hat{s}/k_\perp^2)$ and $I_{12} = 2\frac{1+\beta^2}{2\beta}\ln\frac{1+\beta}{1-\beta}$. Substituting these five integrals into $S_{\mathrm{eik}}$ and applying color conservation $\mathbf{T}_a + \mathbf{T}_b + \mathbf{T}_1 + \mathbf{T}_2 = 0$ collects the diagonal pieces $4\CA\ln(\hat{s}/k_\perp^2) + 4\CF\ln(\hat{s}/m_Q^2) - 4\CF$ into $S_P$ and fixes every coefficient and sign of $\ReP\MS$ in eq.~\eqref{eq:MS_exact} and $R_{\mathrm{EIC}}$ in eq.~\eqref{eq:REIC_exact} uniquely.

\bibliographystyle{JHEP}
\bibliography{references}

\end{document}